\documentclass[journal]{IEEEtran}
\IEEEoverridecommandlockouts
\usepackage{cite}
\usepackage{amsmath,amssymb,amsfonts}
\usepackage{algorithmic}
\usepackage{graphicx}
\usepackage{textcomp}
\usepackage{tabularx}

\usepackage{comment}
\usepackage{url}
\usepackage{balance}
\usepackage[ruled,vlined,linesnumbered]{algorithm2e}
\usepackage{amsmath}
\usepackage{booktabs}
\usepackage{amsfonts}
\usepackage{microtype}  
\usepackage[list=true]{subcaption}
\usepackage[font={footnotesize}]{caption}
\usepackage{times}
\usepackage{multirow}
\usepackage[none]{hyphenat}
\usepackage{float}
\usepackage{acro}
\usepackage{tikz}
\usepackage{pgfplots}
\usetikzlibrary{pgfplots.polar}
\usetikzlibrary{svg.path}
\tikzset{every picture/.style={line width=0.6pt}}
\usetikzlibrary{calc,positioning}
\usetikzlibrary{arrows.meta,
                backgrounds,
                chains,
                fit,
                quotes}
\usepackage[bottom]{footmisc}
\usepackage{xcolor}
\newcommand{\Rev}{\textcolor{black}}
\newcommand{\RevMaj}{\textcolor{black}}

\DeclareAcronym{td}{
short=TD,
long= time domain,
}
\DeclareAcronym{sll}{
short=SLL,
long= sidelobe level,
}
\DeclareAcronym{fd}{
short=FD,
long= frequency domain,
}
\DeclareAcronym{los}{
short=LOS,
long= line-of-sight,
}
\DeclareAcronym{osr}{
short=OSR,
long= over-sampling rate,
}
\DeclareAcronym{5g}{
short=5G,
long= fifth generation,
}
\DeclareAcronym{mu}{
short=MU,
long= multi-user,
}
\DeclareAcronym{cnn}{
short=CNN,
long= convolutional neural network,
}
\DeclareAcronym{imd}{
short=IMD,
long= inter-modulation distortion,
}
\DeclareAcronym{csi}{
short=CSI,
long= channel state information,
}
\DeclareAcronym{zf}{
short=ZF,
long= zero-forcing,
}
\DeclareAcronym{ici}{
short=ICI,
long= intercarrier interference,
}
\DeclareAcronym{dft}{
short=DFT,
long= discrete Fourier transform,
}
\DeclareAcronym{fft}{
short=FFT,
long= fast Fourier transform,
}
\DeclareAcronym{2d}{
short=2D,
long= two-dimensional,
}
\DeclareAcronym{idft}{
short=IDFT,
long= inverse discrete Fourier transform,
}
\DeclareAcronym{bs}{
short=BS,
long= base station,
}
\DeclareAcronym{ue}{
short=UE,
long= user equipment,
}
\DeclareAcronym{iq}{
short=I/Q,
long= quadrature,
}
\DeclareAcronym{I}{
short=I,
long= in-phase,
}
\DeclareAcronym{Q}{
short=Q,
long= quadrature,
}
\DeclareAcronym{ls}{
short=LS,
long= least squares,
}
\DeclareAcronym{ota}{
short=OTA,
long= over-the-air,
}
\DeclareAcronym{sgd}{
short=SGD,
long= stochastic gradient descent,
}
\DeclareAcronym{ce}{
short=CE,
long= cross-entropy,
}
\DeclareAcronym{sl}{
short=SL,
long= supervised learning,
}
\DeclareAcronym{rl}{
short=RL,
long= reinforcement learning,
}
\DeclareAcronym{awgn}{
short=AWGN,
long= additive white Gaussian noise,
}
\DeclareAcronym{ser}{
short=SER,
long= symbol error rate,
}
\DeclareAcronym{qam}{
short=QAM,
long= quadrature amplitude modulation,
}
\DeclareAcronym{rrc}{
short=RRC,
long= root-raised cosine,
}
\DeclareAcronym{snr}{
short=SNR,
long= signal-to-noise ratio,
}
\DeclareAcronym{sinr}{
short=SINR,
long= signal-to-interference-and-noise ratio,
}
\DeclareAcronym{rvftdnn}{
short=RVFTDNN,
long= real-valued focused time-delay neural network,
}
\DeclareAcronym{lo}{
short=LO,
long= local oscillator,
}
\DeclareAcronym{lpf}{
short=LPF,
long= lowpass filter,
}
\DeclareAcronym{pdf}{
short=PDF,
long= probability density function,
}
\DeclareAcronym{cdf}{
short=CDF,
long= cumulative distribution function ,
}
\DeclareAcronym{ccdf}{
short=CCDF,
long= complementary cumulative distribution function ,
}
\DeclareAcronym{fir}{
short=FIR,
long= finite impulse response,
}
\DeclareAcronym{rhs}{
short=RHS,
long= right-hand side,
}
\DeclareAcronym{dsp}{
short=DSP,
long= digital signal processing,
}
\DeclareAcronym{nn}{
short=NN,
long= neural network,
}
\DeclareAcronym{mlp}{
short=MLP,
long=multilayer perceptron
}
\DeclareAcronym{GaN}{
short=GaN,
long=Gallium Nitride,
}
\DeclareAcronym{relu}{
short=ReLU,
long = rectified linear unit, 
}
\DeclareAcronym{gpu}{
short=GPU,
long =graphics processing unit, 
}
\DeclareAcronym{fpga}{
short=FPGA,
long =field-programmable gate array, 
}
\DeclareAcronym{asic}{
short=ASIC,
long =application-specific integrated circuit, 
}
\DeclareAcronym{mse}{
short=MSE,
long=mean squared error,
}
\DeclareAcronym{rvtdnn}{
short=RVTDNN,
long= real-valued time-delay neural network,
}
\DeclareAcronym{arvtdnn}{
short=ARVTDNN,
long= augmented real-valued time-delay neural network,
}
\DeclareAcronym{arden}{
short=ARDEN,
long= attention residual real-valued time-delay neural network,
}
\DeclareAcronym{r2tdnn}{
short=R2TDNN,
long= residual real-valued time-delay neural network,
}
\DeclareAcronym{flop}{
short=FLOP,
long= floating point operations,
}
\DeclareAcronym{ph}{
short=PH,
long= parallel Hammerstein
}
\DeclareAcronym{ofdm}{
short=OFDM,
long=orthogonal frequency division multiplexing,
}
\DeclareAcronym{par}{
short=PAR,
long=peak-to-average ratio,
}
\DeclareAcronym{papr}{
short=PAPR,
long=peak-to-average power ratio,
}
\DeclareAcronym{rf}{
short=RF,
long=radio frequency,
}
\DeclareAcronym{pa}{
short=PA,
long=power amplifier,
}
\DeclareAcronym{pas}{
short=\acs{pa}s,
long=power amplifiers,
}
\DeclareAcronym{psd}{
short=PSD,
long= power spectral densitie,
}
\DeclareAcronym{dpd}{
short=DPD,
long=digital predistortion,
}
\DeclareAcronym{fr1}{
short=FR1,
long=frequency range 1,
}
\DeclareAcronym{fr2}{
short=FR2,
long=frequency range 2,
}
\DeclareAcronym{cfr}{
short=CFR,
long=crest factor reduction,
}
\DeclareAcronym{cf}{
short=CF,
long=crest-factor}

\DeclareAcronym{evm}{
short=EVM,
long=error vector magnitude,
}
\DeclareAcronym{nmse}{
short=NMSE,
long=normalized mean squared error,
}
\DeclareAcronym{trp}{
short=TRP,
long=total radiated power,
}
\DeclareAcronym{oob}{
short=OOB,
long=out-of-band,
}
\DeclareAcronym{acpr}{
short=ACPR,
long=adjacent channel power ratio,
}
\DeclareAcronym{aclr}{
short=ACLR,
long=adjacent channel leakage ratio,
}
\DeclareAcronym{pae}{
short=PAE,
long=power added efficiency,
}
\DeclareAcronym{dla}{
short=DLA,
long=direct learning architecture,
}
\DeclareAcronym{ila}{
short=ILA,
long=indirect learning architecture,
}
\DeclareAcronym{ilc}{
short=ILC,
long=iterative learning control ,
}
\DeclareAcronym{cfr-dpd}{
short=CFR-DPD,
long=CFR combined with DPD,
}
\DeclareAcronym{icf}{
short=ICF,
long=iterative clipping and filtering,
}
\DeclareAcronym{am/am}{
short=AM/AM,
long=amplitude-to-amplitude,
}
\DeclareAcronym{am/pm}{
short=AM/PM,
long=amplitude-to-phase,
}
\DeclareAcronym{siso}{
short=SISO,
long=single-input single-output
}
\DeclareAcronym{mimo}{
short=MIMO,
long=multiple-input multiple-output
}
\DeclareAcronym{mp}{
short=MP,
long=memory polynomial
}
\DeclareAcronym{gmp}{
short=GMP,
long=generalized memory polynomial
}
\DeclareAcronym{adc}{
short=ADC,
long= analog-to-digital converter}
\DeclareAcronym{dac}{
short=DAC,
long= digital-to-analog converter}
\DeclareAcronym{ilc-dpd}{
short=ILC-DPD,
long= adaptive ILC-based DPD
}
\DeclareAcronym{rms}{
short=RMS,
long= root mean squares
}
\DeclareAcronym{vst}{
short=VST,
long= vector signal transceiver
}
\DeclareAcronym{mmwv}{
short=mm-Wave,
long= millimeter-wave
}
\DeclareAcronym{mrt}{
short=MRT,
long= maximum-ratio transmission
}
\DeclareAcronym{toa}{
short=ToA,
long= time-of-arrival
}
\DeclareAcronym{aod}{
short=AoD,
long= angle-of-departure
}

\begin{document}
\bstctlcite{IEEEexample:BSTcontrol}

\title{Time vs. Frequency Domain DPD for Massive MIMO: Methods and Performance Analysis
\thanks{
A limited subset of initial results was presented at  IEEE GLOBECOM 2022, Rio de Janeiro, Brazil~\cite{yibo2022fdcnn}. 

Y. Wu is with Ericsson Research and Chalmers University of Technology, Gothenburg, Sweden (email: yibo.wu@ericsson.com).

U. Gustavsson is with Ericsson Research, Gothenburg, Sweden (e-mail: ulf.gustavsson@ericsson.com).

M. Valkama is with Tampere University, Tampere, Finland (email: mikko.valkama@tuni.fi)

A. Graell i Amat and H. Wymeersch are with Chalmers University of Technology, Gothenburg, Sweden (emails: alexandre.graell@chalmers.se; henkw@chalmers.se).

This work was supported by the Swedish Foundation for Strategic Research (SSF), grant no. ID19-0021, and the Swedish Research Council (VR grant 2022-03007).}}

\author{Yibo~Wu,~\IEEEmembership{Student~Member,~IEEE},
        Ulf~Gustavsson, 
        Mikko~Valkama,~\IEEEmembership{Fellow,~IEEE}, Alexandre~Graell~i~Amat,~\IEEEmembership{Senior~Member,~IEEE}, and
Henk~Wymeersch,~\IEEEmembership{Fellow,~IEEE}
        }

\maketitle

\begin{abstract}
The use of up to  hundreds of antennas in massive \ac{mu} \ac{mimo} \ac{ofdm} poses a  complexity challenge for \ac{dpd} aiming to linearize the nonlinear \acp{pa}. While the complexity for conventional \ac{td} DPD scales  with the number of \acp{pa}, \ac{fd} DPD 
has a complexity scaling with the number of \acp{ue}. In this work, we provide  a comprehensive analysis of different state-of-the-art TD and FD-DPD schemes in terms of complexity and linearization performance in both rich scattering and \ac{los} channels \Rev{and with antenna crosstalk}.  
We propose a novel low-complexity FD \ac{cnn} DPD. \Rev{We also propose a learning algorithm for any FD-DPDs with differentiable structure.} 
The analysis shows that FD-DPD, particularly the proposed FD CNN, is preferable in  LOS scenarios with few users, due to the favorable trade-off between complexity and linearization performance. On the other hand, in scenarios with more users or isotropic scattering channels, significant intermodulation distortions among UEs degrade FD-DPD performance, making TD-DPD more suitable. \Rev{The proposed learning algorithm allows FD-DPDs to outperform TD-DPD optimized by \acl{ila} under antenna crosstalk.}
\end{abstract}

%
\acresetall 
\vspace{-4mm}\section{Introduction}
In 5G and beyond, enhancing the power efficiency of the \acf{pa} is imperative because operating at a high power level often leads to nonlinear distortion~\cite{fager2019linearity}. To operate efficiently, 
\acf{dpd} is a common approach employed to linearize the \ac{pa}~\cite{abdelaziz2018digital}. However, handling up to several hundreds of PAs in massive \ac{mu}-\ac{mimo} systems presents a  complexity challenge for the \ac{dpd}, 
creating a power consumption imbalance that shifts heavily toward DPD~\cite{fager2019linearity}.  Furthermore, DPD typically operates with oversampling rates of five times the signal bandwidth to prevent aliasing effects, which further
exacerbates the power consumption of DPD with high bandwidth signals. \RevMaj{Additionally, low-complexity DPD is crucial for future systems adopting large-scale MIMO technology, such as satellite MIMO and THz MIMO systems~\cite{2024Wu_Satell,2023Wu_THz}, where the wide bandwidth and large number of antennas and PAs present challenges in maintaining high performance and efficiency without excessive power consumption and complexity.} These factors 
motivate the growing need for low-complexity DPD. 

\RevMaj{In fully digital massive MIMO systems, DPD is typically implemented in the \ac{td}, where each PA requires an independent DPD, referred to in this paper as TD-DPD. This approach results in a linear increase in computational complexity with the number of PAs, significantly increasing power consumption and affecting the deployment of digital precoding in \ac{fr1}~\cite{fager2019linearity}. In contrast to hybrid or analog MIMO systems, which utilize fewer \ac{rf} chains than the number of antennas and can share DPD across subarrays or even the entire array~\cite{shared_dpd,BO_DPD_Hybrid,abdelaziz2018digital,brihuega2021frequency}, fully digital MIMO systems cannot take advantage of this power-saving DPD deployment. We consider a fully digital MIMO system operating within \ac{fr1}, with an antenna number scaling up to $100$, as demonstrated in previous testbeds~\cite{vieira2014flexible,malkowsky2017world}.} As an alternative to TD-DPD, placing DPD in the \ac{fd} before the digital precoder and \ac{idft} of OFDM offers a cost-effective solution, known as FD-DPD or beam-domain DPD~\cite{yibo2022fdcnn,brihuega2022beam}. This approach notably reduces complexity, scaling with the number of \acp{ue} rather than PAs, thus addressing the DPD complexity issue. While 
FD-DPD generally has worse linearization performance than TD-DPD, due to its operation in lower dimensions~\cite{yibo2022fdcnn}, higher levels of \acf{oob} nonlinear distortion can be tolerated in \ac{mimo} systems~\cite{mollen2018spatial,mollen2018out}. This makes the choice of TD-DPD vs FD-DPD far from obvious.  


Several works~\cite{brihuega2021frequency,brihuega2022beam,tarver2021virtual, tarver2021ofdm, yibo2022fdcnn} conduct a comparison between FD-DPD and TD-DPD, but only give results in a \ac{los} channel or with an analog-/hybrid- beamforming array architecture. Our initial work~\cite{yibo2022fdcnn} makes a performance and complexity comparison in a Rayleigh fading channel with a fully digital array, albeit focusing solely on in-band linearization performance. In \cite{brihuega2021frequency}, the proposed DPD scheme and comparison are focused exclusively on a single UE and analog-beamforming. \cite{tarver2021virtual,tarver2021ofdm} explores fully-digital MIMO but is limited to frequency-flat \ac{los} channels, with no in-band linearization results for comparison. \cite{brihuega2022beam} considers fully digital MIMO, but it employs a two-stage precoding strategy based on correlated channels, restricting the applicability of the proposed DPD scheme to fully digital precoding. \Rev{ The referenced studies lack comparisons in rich scattering fading channels and ignore the effects of antenna crosstalk, which can be severe due to closely spaced antennas. From the above studies, no clear conclusion can be drawn on which type of DPD is preferable under different channel conditions and antenna crosstalk.} 

\RevMaj{ In this paper, we address the existing knowledge gap in DPD selection for digital MIMO systems by providing a comprehensive analysis of state-of-the-art \ac{td} and \ac{fd} \ac{dpd} schemes, along with a novel \ac{cnn}-based \ac{fd}-\ac{dpd}, for fully digital massive \ac{mu}-\ac{mimo}-\ac{ofdm} systems. Our analysis covers both complexity and performance aspects across various channel scenarios and antenna crosstalk conditions, offering valuable insights for selecting the most suitable \ac{dpd} scheme based on specific computational constraints and linearization requirements.} This work extends our previous study~\cite{yibo2022fdcnn} by generalizing the results to two realistic channel scenarios and introducing new results on \ac{oob} linearization performance under both linear and nonlinear antenna crosstalk. Additionally, we propose a new \ac{fd}-\ac{cnn} structure that further reduces complexity in \ac{mu} scenarios.
Our contributions are: 
\begin{itemize}
    \item \textbf{A new low-complexity FD-DPD:} \RevMaj{We propose a CNN-based DPD called FD-CNN, which addresses the complexity challenges associated with the increasing number of PAs in fully digital massive MIMO systems. Incorporating CNN in the \ac{fd} is novel compared to existing \ac{td} CNN-based DPDs~\cite{liu2019digital,hu2021convolutional}. Moreover, FD-CNN distinguishes itself from existing FD-DPDs by eliminating the need for high-complexity IDFT and DFT operations. As a result, FD-CNN significantly reduces computational complexity compared to existing TD and FD-DPD schemes while meeting the necessary in-band and out-of-band linearization requirements across various scenarios. }
    \item \Rev{\textbf{A new FD-DPD learning algorithm:}
    \RevMaj{We propose a novel FD-DPD learning algorithm applicable to any differentiable FD-DPD model. Training FD-DPDs is challenging because, unlike TD-DPDs—which can be learned using the \ac{ila}, the desired output in FD-DPD is unknown due to domain and dimensionality mismatches. Our algorithm overcomes this issue by enabling supervised learning through pre-trained PA models and precodings. This approach effectively optimizes state-of-the-art FD-DPDs to correct both PA nonlinearity and antenna crosstalk.}
    }
    
    \item \textbf{Comprehensive complexity analysis:}
    \Rev{We conduct a comprehensive complexity comparison between the proposed FD-CNN DPD and other state-of-the-art TD and FD-DPD schemes using the number of FLOPs as a function of the number of BS antennas and UEs \RevMaj{in fully digital massive MIMO systems}. This exact complexity calculation is missing in the literature. This detailed analysis accurately reflects the complexity across different configurations, aiding in the selection of an appropriate DPD within given computational constraints.}
    \item \textbf{Comprehensive performance analysis:}
     \Rev{We compare in-band and OOB linearization performance for various TD and FD-DPD schemes in isotropic scattering and LOS channels, including scenarios with antenna crosstalk, for varying numbers of UEs \RevMaj{in fully digital massive MIMO systems}. This comparison aids in selecting the appropriate DPD for specific linearization needs.  The results show FD-DPD’s effectiveness in scenarios with  few UEs ($<$ 4 in LOS and $<$ 10 in isotropic scattering channels) and under antenna crosstalk, while TD-DPD is more effective in other scenarios.}
    
    
\end{itemize}

The remainder of the paper is organized as follows. Section II outlines the system model. Section III examines PA distortion radiation in two different channel scenarios, motivating the relaxation of \ac{oob} linearization requirements.  Section IV introduces the proposed FD-CNN DPD and contrasts it against commonly used TD and FD DPD schemes, from complexity and performance perspectives. Finally, Section V 
presents a quantitative analysis of the complexity and performance of various DPD schemes across two channel scenarios.

\textit{Notation:} Lowercase and uppercase boldface letters denote column vectors and matrices such as $\boldsymbol{x}$ and $\boldsymbol{X}$; Letters with and without check marks such as  $\check{\boldsymbol{x}}$ and ${\boldsymbol{x}}$ indicate quantities in the \ac{fd} and \ac{td}, respectively.  $\boldsymbol{x}^{\mathsf{H}}$ denotes the Hermitian transpose of $\boldsymbol{x}$;  $\mathbb{R}$ and $\mathbb{C}$ denote real and complex numbers, respectively;   $\boldsymbol{x}[n]$ denote the $n$-th element of $\boldsymbol{x}$, and $\boldsymbol{x}[n:n+k]$ denotes a vector consisting of the $n$-th to $(n+k)$-th elements of $\boldsymbol{x}$; the $B\times U$ all-zeros matrix and the $U\times U$ identity matrix are denoted by $\boldsymbol{0}_{B\times U}$ and $\boldsymbol{I}_{U}$, respectively; $\mathbb{E}_{\boldsymbol{}}\{x\}$  denotes the expectation of $x$.

\begin{figure*}[t]
    \centering
    	\vspace*{-0.5 \baselineskip}
    \includegraphics[width=0.95\linewidth]{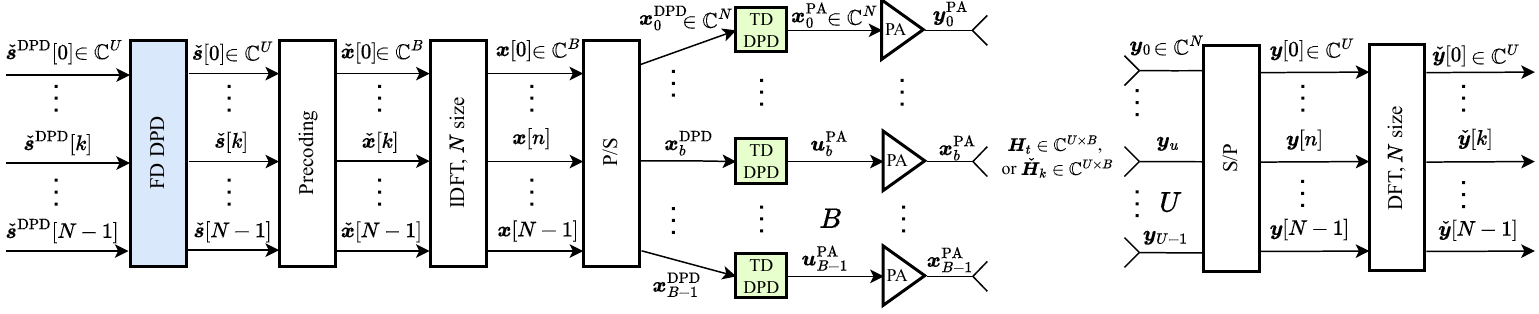}
    \caption{ Baseband equivalent system model of a massive MU-MIMO-OFDM downlink with nonlinear PAs in each RF chain of the BS. The conventional TD per-antenna DPDs take place before each PA while FD-DPD operates before the precoder. \Rev{Crosstalk between each RF chain due to coupling happens before and after the PAs.} }
    	\vspace*{-0.4cm}
    \label{fig:sys_model_MUMIMO}
\end{figure*}
\vspace{-4mm}\section{System Model}
In this section, we provide the basic models of the communication signal, the channel, and the power amplifier. 

\vspace{-4mm}\subsection{Linear MU-MIMO-OFDM System Model}
We consider a massive MU-MIMO-OFDM downlink system model as shown in Fig.~\ref{fig:sys_model_MUMIMO}. The \ac{bs} is equipped with $B$ antennas and transmits messages to $U \ll B$ single-antenna \acp{ue} in the same time-frequency resource using spatial multiplexing. All hardware components, including \ac{dac}, mixer, oscillator, and \acp{pa} are assumed to be ideal.
We use the same OFDM system model as in~\cite{jacobsson2019linear}.  Each OFDM symbol consists of $N$ subcarriers with $N_{\text{d}}$ data subcarriers and $N_{\text{g}}=N-N_{\text{d}}$ guard subcarriers. The subcarrier spacing is denoted by $\Delta f$. Accordingly, the sampling and symbol rate are defined as $f_{\text{s}}=N\Delta f$ and $f_{\text{d}}=
N_{\text{d}}\Delta f$, respectively. The symbol time is denoted by $T_{\text{d}} = 1/f_{\text{d}}$. The upsampling operation is conducted by a large IDFT size $N$ with an \ac{osr} $R = N/N_{\text{d}}$.

\subsubsection{Transmit Signal}
Let $\check{\boldsymbol{s}}[k] \in \mathbb{C}^{U}$ denote the symbol vector for $U$ \acp{ue} at subcarrier $k$ in the \ac{fd}, 
generated independently from a normalized $M$-QAM constellation with $\mathbb{E}\{|\check{{s}}_u[k]|^2\}=1$. Under a flat-fading channel for each subcarrier, the linear precoder maps $\check{\boldsymbol{s}}[k]$ to $\check{\boldsymbol{x}}[k] \in \mathbb{C}^{B}$ as~\cite{LinPrecod_survey,jacobsson2019linear}
\begin{align}
    \check{\boldsymbol{x}}[k] =  \check{\boldsymbol{W}}[k] \check{\boldsymbol{s}}[k], \label{eq:precod_FD}
\end{align}
where $\check{\boldsymbol{W}}[k] \in \mathbb{C}^{B\times U}$ denotes the \ac{fd} precoding matrix for subcarrier $k$. The transmitted signal satisfy the average power constrain $\mathbb{E}\{\| \check{\boldsymbol{x}}[k]\|^2 \}=P_{\text{T}}$. The expectation is over the symbols of all UEs, and $P_T$ denotes the average transmit power, excluding the power amplification by the PAs. In this paper, we consider linear precoders due to their low complexity and good performance~\cite{wiesel2008zero}. 
The precoded vectors $\check{\boldsymbol{x}}[k] $ are transformed to \ac{td} by $N$-size IDFTs, where the \ac{td} signal vector at time sample $n$, $\boldsymbol{x}[n] \in \mathbb{C}^{B}$, is given by
\begin{align}
    \boldsymbol{x}[n]=\frac{1}{\sqrt{N}}\sum_{k=0}^{N-1} \check{\boldsymbol{x}}[k] \exp\left(j2\pi kn/N\right)\,. 
    \label{eq:IDFT_precod}
\end{align}

\subsubsection{Channel Model}
Assuming receivers are much farther from the array than the transmitter's aperture, i.e., in the far-field~\cite{tse2005fundamentals}, the channel's frequency response from the $b$-th antenna of the BS linear array to position $\boldsymbol{p} \in \mathbb{R}^3$ at subcarrier $k$ is modeled as~\cite{clerckx2013mimo}
\begin{align}
    \check{{h}}_{\boldsymbol{p},b}[k] = \frac{1}{\sqrt{L}}\sum_{l=0}^{L-1} \beta_{\boldsymbol{p},l} e^{-j2\pi (f_c \tau_{\boldsymbol{p},l}+b\sin{\theta_{\boldsymbol{p},l}}/2)}  e^{-j2\pi k\Delta f \tau_{\boldsymbol{p},l}},\label{eq:channel}
\end{align}
where $\tau_{\boldsymbol{p},l}$ and $\theta_{\boldsymbol{p},l}$ are, respectively, the delay and the \ac{aod} of the signal from the linear array to position $\boldsymbol{p}$ associated with  path $l$. $f_c$ is the carrier frequency. In total, there are $L$ propagation paths with uniformly distributed delays between $0$ and the delay spread $\delta_{\tau}$. The number of significant taps in the channel response is defined by $\gamma=\lceil R \sigma_{\tau}/T_{\text{d}}\rceil$. The large-scale fading, including the pathloss and shadowing, from the array to position $\boldsymbol{p}$ for the $t$-th path is modeled by  $\beta_{\boldsymbol{p},l} \in \mathbb{R}^{+}$. 

We use the channel response~\eqref{eq:channel} to model two channel scenarios: frequency-selective isotropic fading and frequency-flat \ac{los} propagation. 
\begin{itemize}
    \item 
\emph{Frequency-selective isotropic fading:} The channel response \eqref{eq:channel} models isotropic fading by assuming that the number of paths $L$ is large  with \ac{aod} uniformly distributed over $[-\pi/2,\pi/2]$. Then it is common to use the uncorrelated Rayleigh fading channel model~\cite{clerckx2013mimo}
\begin{align}
     \check{{h}}_{\boldsymbol{p},b}[k] = \frac{1} {\sqrt{\gamma}}\sum_{t=0}^{\gamma-1}  h_{\boldsymbol{p},b,t} e^{-j2\pi k t/N},\label{eq:channel_Rayleigh}
\end{align}
where each time domain channel tap gain is independently distributed as  $h_{\boldsymbol{p},b,t}\sim \mathcal{N}_{\mathbb{C}}(0, \sigma_{\beta_{\boldsymbol{p}}}^2)$ with large-scale fading coefficient $ \sigma_{\beta_{\boldsymbol{p}}}^2$.  
Different numbers of channel taps $\gamma$ model different degrees of frequency selectivity.  
\item 
\emph{Frequency-flat \ac{los} propagation:} The channel response  \eqref{eq:channel} models pure \ac{los} propagation assuming that the number of paths $L=1$ and the channel response is the same for all subcarriers. Thus, we can rewrite the channel response~\eqref{eq:channel} by dropping the path subscript (i.e., $\beta_{\boldsymbol{p},0}=\beta_{\boldsymbol{p}}$ and $\theta_{\boldsymbol{p},0}=\theta_{\boldsymbol{p}}$) as 
\begin{align}
\check{{h}}_{\boldsymbol{p},b}[k] = \beta_{\boldsymbol{p}} e^{-j2\pi (f_c \tau_{\boldsymbol{p}}+b\sin{\theta_{\boldsymbol{p}}/2)}},
    \label{eq:channel_LOS} 
\end{align}
where the \ac{aod} of the \ac{los} path are defined based on the geometry by $\theta_{\boldsymbol{p}} = \arccos{\left({{p}[1] - {p}_{\text{BS}}[1]}/{\| \boldsymbol{p} - \boldsymbol{p}_{\text{BS}}\|}\right)}$. Here $\boldsymbol{p}_{\text{BS}}$ denotes the BS position.
\end{itemize}

The channel frequency response from the array to user $u$ at position $\boldsymbol{p}_u$ and subcarrier $k$ is given  by 
\begin{align}
\check{\boldsymbol{h}}_{\boldsymbol{p}_u}[k]=[\check{{h}}_{\boldsymbol{p}_u,0}[k],\cdots,\check{{h}}_{\boldsymbol{p}_u,B-1}[k]]^{\mathsf{T}} \in \mathbb{C}^{B}. \label{eq:channel_array}
\end{align}
Stacking all $\check{\boldsymbol{h}}_{\boldsymbol{p}_u}[k]$ from different UEs into a matrix yields the channel matrix for all $U$ UEs at subcarrier $k$,
\begin{align}
    \check{\boldsymbol{H}}[k]\triangleq \left[\check{\boldsymbol{h}}_{\boldsymbol{p}_0}[k],\cdots,\check{\boldsymbol{h}}_{\boldsymbol{p}_{U-1}}[k]\right],
\end{align}
where we drop the subscript $\boldsymbol{p}_u$ for notation simplicity.
The corresponding channel matrix in the \ac{td} at time sample $n$ is 
\begin{align}
    \boldsymbol{H}_{}[n] = \frac{1}{\sqrt{N}}\sum_{k=0}^{N-1} \check{\boldsymbol{H}}_{}[k] \exp(j2\pi kn/N),
\end{align}
where the time sample index $n \in \{0,1,\cdots,\gamma-1\}$.

\subsubsection{Received Signal}
Assuming a linear MU-MIMO (i.e., ignoring for now the  nonlinear \ac{pa} and \ac{dpd}), 
%
the signal vector received at the $U$ UEs, $\boldsymbol{y}_{}[n]\in\mathbb{C}^{U}$, is given by~\cite{bjornson2017massive}
\begin{align}
    \boldsymbol{y}_{}[n] = \sum_{n^{\prime}=0}^{\gamma-1} {\boldsymbol{H}}_{}^{}[n^{\prime}] {\boldsymbol{x}}[n-n^{\prime}] + \boldsymbol{\eta}[n], \label{eq:Rx_at_UEs_TD}
\end{align}
where $\boldsymbol{\eta}[n] \sim  \mathcal{N}_{\mathbb{C}}\left(\mathbf{0}_{U \times 1}, \sigma^2 \mathbf{I}_U\right)$ models \ac{awgn} at the $U$ UEs at time sample $n$ with noise power $\sigma^2$.  Eq. \eqref{eq:Rx_at_UEs_TD} can be written in the \ac{fd} as~\cite{bjornson2017massive}
\begin{align}
    \check{\boldsymbol{y}}_{}[k] = \check{\boldsymbol{H}}_{}^{}[k] \check{\boldsymbol{x}}[k] + \check{\boldsymbol{\eta}}[k], \label{eq:Rx_at_UEs_FD}
\end{align}
where $\check{\boldsymbol{y}}[k] = 1/\sqrt{N}\sum_{n=0}^{N-1} {\boldsymbol{y}}[n] \exp (-j2 \pi k n/N)$ and $\check{\boldsymbol{\eta}}[k] = 1/\sqrt{N}\sum_{n=0}^{N-1} {{\eta}}[n] \exp (-j2 \pi kn/N)$ are the corresponding FD received signal and noise for the $U$ UEs at subcarrier $k$, respectively. 

Substituting~\eqref{eq:precod_FD} into~\eqref{eq:Rx_at_UEs_FD}, $\check{\boldsymbol{y}}[k]$ with linear precoder $\check{\boldsymbol{W}}[k]$ can be expressed as~\cite{wiesel2008zero}
\begin{align}
    \check{\boldsymbol{y}}[k] = \check{\boldsymbol{H}}[k] \check{\boldsymbol{W}}[k] \check{\boldsymbol{s}}[k] + \check{\boldsymbol{\eta}}[k]\,.
    \label{eq:channel_FD_with_precoding}
\end{align}

\subsubsection{Precoding Strategies}

We will consider two commonly used BS linear precoders: \ac{mrt} and \ac{zf} precoding~\cite{wiesel2008zero}. 
The \ac{mrt} precoder maximizes the power directed toward each UE and ignores \ac{mu} interference. The \ac{mrt} precoding matrix in~\eqref{eq:channel_FD_with_precoding} is given by
   $\check{\boldsymbol{W}}^{\text{MRT}}[k] = \alpha^{\text{MRT}} \check{\boldsymbol{H}}^{\mathsf{H}}[k],$
where $\alpha^{\text{MRT}}$ is the normalization factor to ensure the power constraint $\mathbb{E}\left\{\|\check{\boldsymbol{x}}[k]\|^{2}\right\}=P_{T}$ is met~\cite{wiesel2008zero}. 
%
The \ac{zf} precoder minimizes the MU interference by using the pseudo-inverse of the channel matrix is used as the precoding matrix~\cite{wiesel2008zero}. The \ac{zf} precoding matrix $\check{\boldsymbol{W}}[k]$ in~\eqref{eq:channel_FD_with_precoding} is given by~\cite{wiesel2008zero}
\begin{align}
\check{\boldsymbol{W}}^{\text{ZF}}[k] = \alpha^{\text{ZF}} \check{\boldsymbol{H}}^{\mathsf{H}}[k](\check{\boldsymbol{H}}[k] \check{\boldsymbol{H}}^{\mathsf{H}}[k])^{-1},
    \label{eq:Zero_forcing}
\end{align}
where $\alpha^{\text{ZF}}$ denotes the normalization factor to satisfy the same power constraint as the \ac{mrt} precoder.
\vspace{-4mm}\subsection{Nonlinear Power Amplifier in MU-MIMO-OFDM} \label{section:sys_model_PA}
After the IDFT, the \ac{td} OFDM symbols are mapped to $B$ \ac{rf} chains and converted to the analog domain by the \acp{dac} (ignoring the DPD). For simplicity, we assume ideal \acp{dac} with infinite-resolution. The PA input signal at the $b$-th RF chain is $\boldsymbol{u}_b^{\text{PA}} \triangleq \left[x_0[b],\cdots,x_{N-1}[b]\right]^{\mathsf{T}} \in \mathbb{C}^{N}$.\footnote{\vspace*{-0\baselineskip} In Section IV-A, we describe  how the PA input $\boldsymbol{u}_b^{\text{PA}}$ is obtained in the presence of DPD. \vspace*{-0\baselineskip}} 
%
%
Specifically, the PA associated with the $b$-th BS antenna is commonly modeled by the \ac{gmp} model~\cite{GMP_2006}, with nonlinear order $Q^{\text{PA}}$,  memory length $M^{\text{PA}}$, cross-term length $G^{\text{PA}}$, input ${u}_b^{\text{PA}}[n: n-M^{\text{PA}}]=\left[{u}_b^{\text{PA}}[n], \ldots, {u}_b^{\text{PA}}[n-M^{\text{PA}}]\right]^{\mathsf{T}} \in \mathbb{C}^{M^{\text{PA}}+1}$, and output $x^{\text{PA}}_{b,n}$. The input-output relation of the $b$-th GMP-based PA model at time sample $n$ can be expressed as
\begin{equation}
    \begin{aligned}
x^{\text{PA}}_{b}&[n]= \sum_{q=0}^{Q^{\text{PA}}-1} \sum_{m=0}^{M^{\text{PA}}} a_{q,m}^{(b)} {u}_b^{\text{PA}}[n-m]\big| {u}_b^{\text{PA}}[{n-m}]\big|^{q} \\
&+\sum_{q=1}^{Q^{\text{PA}}-1} \sum_{m=0}^{M^{\text{PA}}} \sum_{g=1}^{G^{\text{PA}}}\Big(c_{q,m,g}^{(b)} {u}_b^{\text{PA}}[{n-m}]\big| {u}_b^{\text{PA}}[{n-m-g}]\big|^{q}\\
&+e_{q,m,g}^{(b)} {u}_b^{\text{PA}}[{n-m}]\left|{u}_b^{\text{PA}}[{n-m+g}]\right|^{q}\Big)\,,
\end{aligned}
\label{eq:PA_GMP_b}
\end{equation}
where $a_{q,m}^{(b)}$, $c_{q,m,g}^{(b)}$, and $e_{q,m,g}^{(b)}$ are coefficients for PA $b$. \RevMaj{The GMP model~\eqref{eq:PA_GMP_b} consists of polynomial terms with memory components, divided into three parts. The first part, known as the \ac{mp} model~\cite{ding2004robust}, includes $Q^{\text{PA}}M^{\text{PA}}$ coefficients, $\{a_{q,m}^{(b)}\}$, capturing interactions of the input signal itself. The second part contains lagging cross-terms, with $Q^{\text{PA}}M^{\text{PA}} G^{\text{PA}}$ coefficients, $\{c_{q,m}^{(b)}\}$, capturing interactions between current and previous inputs. The final part consists of leading cross-terms, also with $Q^{\text{PA}}M^{\text{PA}} G^{\text{PA}}$ coefficients, $\{e_{q,m}^{(b)}\}$, capturing interactions between the current and future inputs.}

The average power gain provided by the \acp{pa} is defined as $G=\mathbb{E}\{ |x^{\text{PA}}_{b}[n]|^2  /  |u^{\text{PA}}_{b}[n]|^2 \}$, where the expectation is taken over all samples for all PAs. \RevMaj{With an ideal linear PA, the desired output of the $b$-th PA at time sample $n$ is denoted by $\Acute{x}_b^{\text{PA}}[n]\triangleq G u_b^{\text{PA}}[n]$.}  Because of precoding, the average PA output power of each PA, defined as $P^{\text{PA}}_b = \mathbb{E}\{|x^{\text{PA}}_{b}[n]|^2 \}$ for  the $b$-th PA output, can be different. The average output power of the BS is defined as $P^{\text{BS}} = \sum_{b=0}^{B-1} P^{\text{PA}}_b = GP_{\text{T}}$.

Nonlinear \acp{pa} cause distortion that disrupts subcarrier orthogonality, resulting in \ac{ici}. They also give rise to \ac{mu} interference due to the nonlinear \ac{imd}, even when ideal \ac{zf} precoding is employed. The nonlinear distortion appearing in the \ac{oob} frequencies may interfere with neighboring communication systems in adjacent frequency bands. This OOB distortion for large array can demonstrate directional beamforming effects, which we will further explore in the following section.
\Rev{
\vspace{-4mm}\subsection{Linear and Nonlinear Crosstalk} \label{section:antenna_cross}
In massive MIMO systems, crosstalk arises due to the coupling between closely spaced antennas and signal paths. Crosstalk occurring before the PA leads to nonlinear crosstalk, while crosstalk occurring after the PA results in linear crosstalk~\cite{amin2014behavioral}. In the presence of  nonlinear crosstalk, the input to the $b_1$-th PA with crosstalk, $\tilde{\boldsymbol{u}}_{b_1}^{\text{PA}}$, can be modeled as:
\begin{align}
    \tilde{\boldsymbol{u}}_{b_1}^{\text{PA}} = {\boldsymbol{u}}_{b_1}^{\text{PA}} + \sum\nolimits_{\substack{b_2=0 \\ b_2\neq b_1}}^{B-1} g^{\text{In}}_{b_1,b_2} {\boldsymbol{u}}_{b_2}^{\text{PA}}, 
    \label{eq:nonlinear_crosstalk}
\end{align}
which introduces nonlinear crosstalk when transmitted through each nonlinear PA~\eqref{eq:PA_GMP_b}. Here $g^{\text{In}}_{b_1,b_2} \in \mathbb{C}$ is the crosstalk parameter between the inputs of PA $b_1$ and $b_2$.\footnote{The crosstalk parameter between other pairs of branches diminish with the square of the physical distance, and the crosstalk phases were derived based on the effective physical distance relative to the wavelength.} 
Similarly, in the case of linear crosstalk, the output of the $b_1$-th PA can be modeled as
\begin{align}
    \tilde{\boldsymbol{x}}_{b_1}^{\text{PA}} = {\boldsymbol{x}}_{b_1}^{\text{PA}} + \sum\nolimits_{\substack{b_2=0 \\ b_2\neq b_1}}^{B-1} g^{\text{Out}}_{b_1,b_2} {\boldsymbol{x}}_{b_2}^{\text{PA}}
        \label{eq:linear_crosstalk}
\end{align}
where $g^{\text{Out}}_{b_1,b_2} \in \mathbb{C}$ is the crosstalk parameter between the outputs of PA $b_1$ and $b_2$. We assume memoryless crosstalk as in~\cite{amin2014behavioral,brihuega2020digital}. Crosstalk degrades beamforming performance by causing beams to become less focused or misdirected, reducing array gain. Furthermore, nonlinear crosstalk exacerbates the nonlinear characteristics of the PA, resulting in spectral regrowth and increased emissions both in-band and out-of-band. Consequently, the radiation pattern may exhibit unexpected lobes and sidelobes due to these nonlinear effects.
}

\Rev{
\vspace{-4mm}\section{Relaxation of OOB Linearization Requirements in Massive MIMO Systems}\label{section:PA_distortion_radiation}
Nonlinear \acp{pa} generate distortions that affect both the in-band and \ac{oob} frequencies. In many massive MIMO scenarios, in-band signals receive higher power gain than OOB signals due to beamforming. Thus, MIMO systems can transmit with less power to achieve the same in-band signal power at the UE as a legacy \ac{siso} system, resulting in less received \ac{oob} distortion~\cite{mollen2018spatial}. This relaxes the OOB linearization requirements for DPD. The degree of relaxation depends on channel characteristics, BS antennas, users, and the amount of crosstalk. This section summarizes PA distortion radiation and discusses DPD linearization requirement relaxation for \ac{los} and isotropic  channels. Detailed derivations can be found in~\cite{mollen2018spatial}, though unlike~\cite{mollen2018spatial}, 
we consider PAs with memory.
}
\Rev{
\vspace{-4mm}\subsection{OOB Linearization Relaxation in a Line-of-Sight Channel}\label{section:Dist_pattern_LOS}
In a \ac{los} channel, directive beamforming focuses transmission beams toward the UE direction but can also beamform distortion in certain directions, harming potential receivers. In a frequency-flat \ac{los} channel with MRT precoding, third-order distortion, which often dominates, is beamformed in approximately $U^3$ distinct directions~\cite[Theorem 2]{mollen2018spatial}. In these directions, distortion builds up constructively, while in other directions, it destructively interferes. Maximum power gains are achieved for both linear and distorted signals in the UE directions, with total distortion power at each UE scaling as $B/U^2$~\cite[Remark 3, Theorem 1]{mollen2018spatial}. As $U$ increases, distortion power decreases until it saturates at a level determined by $\beta_{\boldsymbol{p}}$.}

\begin{figure}[t]
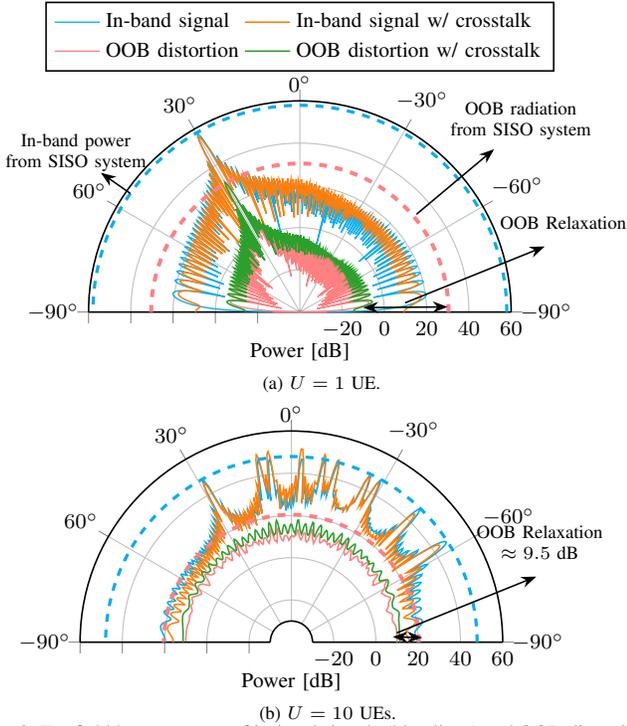

    \begin{subfigure}{0.5\textwidth}
     \input{figures/BP_example_LOS_U1}
    \vspace{-0.15cm}
    \caption{$U=1$ UE.}
    \label{fig:BP_LOS_example_U1}
    \end{subfigure}
\hfill
\vspace{-0.1cm}
\begin{subfigure}{0.5\textwidth}
    \centering
     \input{figures/BP_example_LOS_U10}
        \vspace{-0.15cm}
    \caption{$U=10$ UEs.}
    \label{fig:BP_LOS_example_U10}
\end{subfigure}
\caption{Far-field beampatterns of in-band signals (blue lines) and OOB distortion (red lines) using $B=100$ antennas and $U=\{1,10\}$ UEs in a pure LOS channel, without DPD. Dashed lines represent in-band and OOB radiations from a SISO system, with transmit power calibrated to achieve equivalent in-band power at the UEs as in MIMO systems. }
    \label{fig:BP_LOS_example_all}
\vspace{-2mm}    
\end{figure}
\begin{figure}[t]
    \centering
     \input{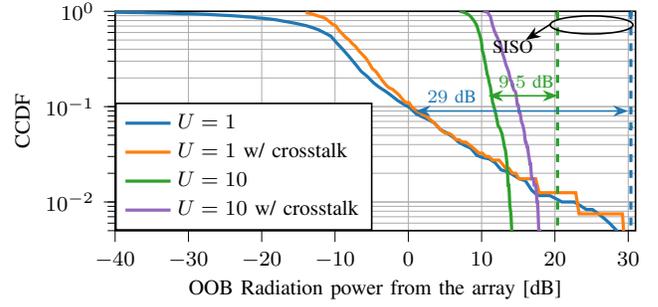}
    \caption{Distributions of the OOB distortion power (results from Fig.~\ref{fig:BP_LOS_example_all})  from a $B=100$ antennas array with varying number of UEs $U=\{1,10\}$ in a  LOS channel, without DPD. Dashed lines represent in-band and OOB radiations from a SISO system, with aligned transmit power to achieve equivalent in-band power at the UEs as in MIMO systems. }
    \label{fig:OOB_distri_LOS}
\end{figure}

\Rev{
This OOB linearization relaxation is illustrated in Fig.~\ref{fig:BP_LOS_example_all}, showing far-field beampatterns for $B=100$ antennas of the in-band signal and OOB distortion under equal power allocation among $U=\{1,10\}$ users.\footnote{\RevMaj{We consider a fully digital MIMO system with 100 antennas, which is a reasonable limit as supported by prior studies~\cite{vieira2014flexible,malkowsky2017world}.}} The results include scenarios with $-10$ dB linear and nonlinear crosstalk, and isotropic beampatterns from a SISO system.
}
\Rev{
Fig.\ref{fig:BP_LOS_example_U1} shows the case for $U=1$. The OOB distortion is beamformed in the same direction as the served UE with nearly the same power gain as the in-band signal, resulting in OOB radiation similar to a SISO system. However, OOB distortion power in other directions is negligible, significantly relaxing OOB requirements in non-UE directions. This relaxation can also apply to the UE direction due to the low probability of an unfortunate victim being in the same direction with a large array (probability $\propto 1/B$). Crosstalk slightly increases both in-band and OOB emissions, affects the beamforming direction, and amplifies some sidelobes. As the number of UEs increases, distortion radiation becomes nearly isotropic with minimal power gain in all directions, leading to uniform OOB relaxation approximately equal to the array gain $B/U$. Fig.~\ref{fig:BP_LOS_example_U10} illustrates this for $U=10$ UEs, showing a relaxation of about $10$ dB. The impact of crosstalk on the power and direction of in-band and output becomes more pronounced compared to the single UE case.
}
\Rev{
Fig.~\ref{fig:OOB_distri_LOS} provides another view, illustrating the distribution of OOB radiation over a uniformly distributed \ac{aod} $\theta_{\boldsymbol{p}} \in [-\pi/2, \pi/2]$. For small $U$, significant OOB relaxation is achievable (e.g., $29$ dB for $U=1$ for $90\%$ of angles). For larger $U$, the CDF curves are more vertical due to less directive OOB radiation, resulting in around $9.5$ dB OOB relaxation for $U=10$. As before, crosstalk increases OOB radiation.
}

\Rev{
\vspace{-7mm}\subsection{OOB Linearization Relaxation in Isotropic Fading Channels}\label{section:PA_distortion_isotropic}
}
\begin{figure}
    \centering
    \input{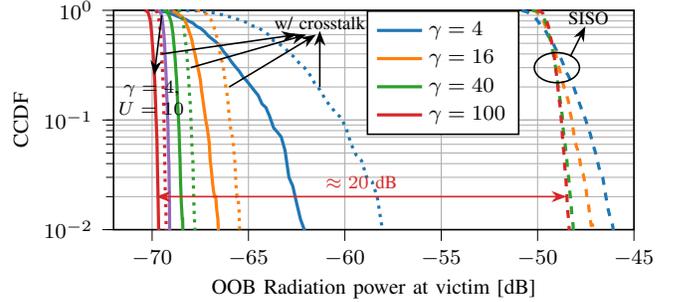}
    \caption{Distributions of the OOB distortion power from a linear array to a victim receiver randomly located  at a fixed distance of $25$ m from the array, without DPD. There are $B=100$ antennas and $U=1$ UEs. The isotropic fading channel is evaluated with different channel taps $\gamma=\{4,16,40,100\}$. The case of $\gamma=4$ and $U=10$ is also shown. Dashed lines represent OOB radiations from SISO systems, with aligned transmit power to achieve equivalent in-band power at the UEs as in MIMO systems. \Rev{Dotted lines represent OOB radiations with crosstalk.}}
    \label{fig:OOB_distri_vs_gamma}
\end{figure}
\Rev{
When the channel~\eqref{eq:channel} is frequency-selective with multiple paths and varied delays (typical of sub-6 GHz bands), frequency-selective beamforming directs the same data stream along each path with corresponding delays. This expands the beamforming directions, making OOB radiation less directive, similar to serving more \acp{ue} in the LOS scenario~\cite{mollen2018out}. As the number of channel taps increases, the OOB radiation power distribution becomes more isotropic~\cite{mollen2018spatial}.
Fig.~\ref{fig:OOB_distri_vs_gamma} illustrates this isotropic relationship for different channel taps $\gamma={4,16,40,100}$, with OOB radiation results for a victim receiver at the same distance as the UE.\footnote{Similar results and conclusions can be obtained as in~\cite{mollen2018spatial} by setting an \ac{osr} $R=4$ and the delay spreads $\sigma_{\tau}=\{1,4,10,25\} T_{\text{d}}$ in the channel model~\eqref{eq:channel} with large paths, uniform power-delay profile, and uniform AoD distribution.} A linear array with $B=100$ antennas and a single UE $U=1$ is considered. SISO system results are shown in dashed lines. Results with the same crosstalk level as in the LOS scenario are shown in dotted lines. The curves become more vertical with increasing channel taps, indicating more isotropic OOB radiation. The rightward shift with fewer channel taps shows increased OOB power due to higher power variations between PAs in different RF chains. Although crosstalk increases OOB radiation, its impact diminishes as $\gamma$ increases, eventually becoming negligible as OOB radiation becomes isotropic.
}
Finally, we also show the case of $U=10$ and $\gamma=4$. The OOB radiation remains relatively constant in the multi-antenna case but is reduced in the SISO case due to user power sharing. This shows that OOB relaxation depends on $B/U$. 

Therefore, we conclude that in a frequency-selective isotropic fading channel, the OOB requirement for DPD can be relaxed due to reduced-power isotropic distortion, similar to the scenario with multiple UEs in LOS channels.


\section{Frequency Domain Convolutional Neural Network DPD}
Since the linearization requirement for DPD can be relaxed in many  MU-MIMO-OFDM scenarios, we can leverage this relaxation to design low-complexity FD DPDs. However, both state-of-the-art TD and FD DPDs suffer from complexity issues. In this section, we briefly analyze state-of-the-art TD and FD DPDs and then 
introduce FD-CNN, a novel FD-DPD model based on \ac{cnn} that eliminates the need for Fourier transforms for FD-DPD, enhancing flexibility.  \Rev{We perform a quantitative complexity analysis, as well as a qualitative analysis in terms of implementation aspects, including the scalability to BS antennas and UEs, different parameter learning methods, adaptability to two types of channels, and ability to handle \ac{imd} products, aspects often missing in the literature. }
\vspace{-4mm}
\subsection{State-of-the art DPD}

\subsubsection{State-of-the-art Time-Domain DPD}
Among the various TD-DPD models, we opt for the widely-used \ac{gmp} model~\cite{GMP_2006}, as shown in~\eqref{eq:PA_GMP_b}, due to its superior balance between linearization performance and computational complexity compared to other TD models, such as Volterra-based and \ac{nn}-based approaches. Thus, the TD-GMP model is a common benchmark for assessing DPD complexity in many papers~\cite{complexityGMP, Joint_IQ_MIMO, AVDTDNN, wu2021low}. \Rev{The input-output relation of the TD GMP-based DPD for the $b$-th RF chain can be constructed by feeding $u_b^{\text{DPD}}[n]$ as the input to the GMP model~\eqref{eq:PA_GMP_b}. Details are omitted for presentation brevity.
}

\subsubsection{State-of-the-art Frequency-Domain DPDs}
\Rev{Compared with TD DPD, few FD DPD schemes exist with limited models for nonlinear PA behavior in the frequency domain. We analyze two recent FD DPD approaches, namely~\cite{brihuega2022beam} and \cite{tarver2021virtual}, which both apply time-domain DPD models in the frequency domain via (I)DFTs. }
\paragraph*{Model-based Approaches}
The \ac{fd}-\ac{mp} \ac{dpd} was initially introduced in~\cite{brihuega2021frequency} for a single UE MIMO with an analog beamforming system. It was later generalized to an MU scenario with digital precoding in~\cite{brihuega2022beam}, achieving complexity reduction through a two-stage digital precoding approach that assumes correlated channels between antennas. 
In Fig.~\ref{fig:sys_model_MUMIMO}, FD DPD is placed before the digital precoding. To align with the TD-GMP, the GMP model is adopted in the FD and referred to as FD-GMP.
\Rev{
In FD-GMP, each UE stream is first transformed into \ac{td} via \ac{idft}. Specifically, given the symbol vector for the $u$-th \ac{ue} in the \ac{fd}, $\tilde{\boldsymbol{s}}_u^{\text{DPD}} \in \mathbb{C}^{N} = \boldsymbol{F}_N^{\mathsf{H}}\tilde{\boldsymbol{s}}_u^{\text{DPD}}$
where $\boldsymbol{F}_N$ denotes the unitary $N$-size DFT matrix. This TD symbol vector is then fed to a GMP model~\eqref{eq:PA_GMP_b}~\cite{brihuega2021frequency}. Details are omitted for presentation brevity. The FD-GMP output ${\boldsymbol{s}}_{u}\in \mathbb{C}^{N}$ is transformed back into FD by DFT: $\check{\boldsymbol{s}}_{u}= \boldsymbol{F}_N {\boldsymbol{s}}_{u}\in \mathbb{C}^{N}$. 
}\Rev{
To cancel out each \ac{imd} beam between UEs, a separate FD-GMP model can be utilized, which is why FD-GMP is also known as \textit{beam-domain} DPD. 
}
 \paragraph*{NN-based Approaches}
An alternative  \ac{nn} -based DPD in the FD, introduced in~\cite{tarver2021virtual} for massive MU-MIMO, operates before the precoder. We refer to this model as FD-NN. Similar to FD-GMP, each \ac{ue} stream is first transformed into \ac{td} via \acp{idft}. Followed by the construction of the \ac{nn} input, the UE signal with memory is then fed into the  \ac{nn}, where
all layers are fully connected. The FD-NN has $K^{\text{NN}}$ hidden layers, each with $D$ neurons and a ReLU nonlinear activation function. The output layer has $2U$ neurons. Similar to FD-GMP, the predistorted \ac{td} signal for each UE is transformed back to the \ac{fd} by \acp{dft} before being sent to the digital precoder.

\subsection{Proposed \ac{fd}-CNN: Architecture} \label{section:FD_CNN}
 \begin{figure*}[t!]
    \centering
    	\vspace*{0 \baselineskip}
    \includegraphics[width=0.9\linewidth]{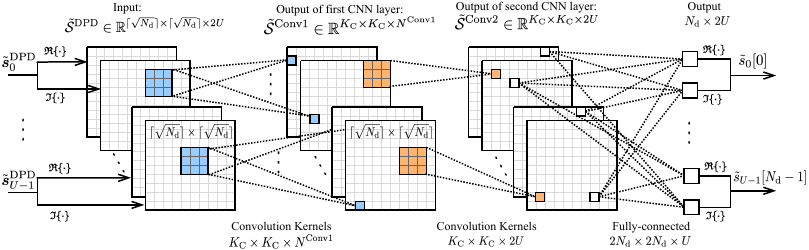}
    \caption{Structure of the proposed FD-CNN. The input of FD-CNN is formed by the $U$ UE symbol vectors, $[\tilde{\boldsymbol{s}}_0^{\text{DPD}},\cdots \tilde{\boldsymbol{s}}_{U-1}^{\text{DPD}}]$. The output is the predistorted symbol vectors of the $U$ UEs, $[\tilde{\boldsymbol{s}}_{0},\cdots,\tilde{\boldsymbol{s}}_{U-1}]$.}
    \label{fig:FD-CNN-diagram}
    	\vspace*{-0.5cm}
\end{figure*}
To address the additional precoding cost introduced by (I)DFTs in both FD-GMP and FD-NN, the proposed FD-CNN leverages the \ac{cnn}. The structure of the proposed FD-CNN is depicted in Fig.~\ref{fig:FD-CNN-diagram}. FD-CNN operates in the \ac{fd} preceding the digital precoder, taking UE symbols as input. It comprises three \ac{nn} layers, each serving a distinct function. 
\subsubsection{Input Layer}
The first layer consists of $N^{\text{Conv1}}$ \ac{2d} convolutional filters, designed to efficiently extract relevant information from the \ac{fd} UE symbol vectors. This functionality serves a similar purpose to the IDFTs in FD-NN~\cite{tarver2021virtual} and FD-GMP~\cite{brihuega2021frequency,brihuega2022beam}. The utilization of \ac{2d} convolutional layers offers computational advantages compared to employing fully connected layers because the input size scales linearly with the number of subcarriers. To facilitate the 2D convolution, the \ac{fd} UE symbol vector $\tilde{\boldsymbol{s}}_u^{\text{DPD}} \in \mathbb{C}^{N}$ is converted to a matrix $\tilde{\boldsymbol{S}}_u^{\text{DPD}} \in \mathbb{C}^{\lceil\sqrt{N_{\text{d}}}\rceil \times \lceil\sqrt{N_{\text{d}}} \rceil}$,\footnote{\vspace*{0\baselineskip} This vector-to-matrix conversion follows a contiguous order, ensuring that convolution kernels can effectively extract information from adjacent subcarriers. \vspace*{-0\baselineskip}} where only $N_{\text{d}}$ data subcarriers are involved to save complexity. Here $\lceil\cdot\rceil$ represents the ceiling function, and zero-padding is applied if $\lceil\sqrt{N_{\text{d}}} \rceil> \sqrt{N_{\text{d}}}$. All $U$ UE symbol vectors are converted to matrices, collectively forming the input of the FD-CNN as, 
\begin{align}
	\tilde{\mathcal{S}}^{\text{DPD}} = [\tilde{\boldsymbol{S}}_{0}^{\text{DPD}},...,\tilde{\boldsymbol{S}}_{{u}}^{\text{DPD}},...,\tilde{\boldsymbol{S}}_{{U-1}}^{\text{DPD}} ].
	\label{eq:FD-CNN-input}
\end{align}
To provide an intuitive illustration of this process, envision that the $U$ UE symbol vectors are transformed into $U$ images, each of which undergoes feature extraction via numerous 2D convolutional filters. 

To use real-valued  \ac{nn}s, each complex-valued UE symbol matrix $\tilde{\boldsymbol{S}}_{{u}}^{\text{DPD}}$ is decomposed into real and imaginary matrices with the same dimension. This results in $2U$ real-valued UE symbol matrices, which are convoluted with $N^{\text{Conv1}}$ convolutional kernels. Consequently, the output tensor of the first layer, denoted as $\tilde{\mathcal{S}}_{}^{\text{Conv1}} \in \mathbb{R}^{\lceil\frac{\lceil\sqrt{N_{\text{d}}}\rceil}{K_{\text{S}}}\rceil \times \lceil\frac{\lceil\sqrt{N_{\text{d}}}\rceil}{K_{\text{S}}}\rceil \times N^{\text{Conv1}}}$, comprises $N^{\text{Conv1}}$ matrices $\{\tilde{\boldsymbol{S}}_{\omega}^{\text{Conv1}}, \omega=0,\cdots,N^{\text{Conv1}}-1\}$. The $\omega$-th output matrix $\tilde{\boldsymbol{S}}_{\omega}^{\text{Conv1}}$ is obtained by convolving with the $\omega$-th convolutional kernel, as follows:
\begin{align}
    \tilde{\boldsymbol{S}}_{\omega}^{\text{Conv1}} = f^{\sigma}(f^{\text{Conv1}}_{\omega}(\tilde{\mathcal{S}}^{\text{DPD}})),\label{eq:FDCNN_Conv1}
\end{align}
where $f^{\sigma}(\cdot)$ denotes an element-wise activation function, and $f^{\text{Conv1}}_{\omega}(\cdot)$ represents the $\omega$-th convolutional kernel with a kernel size $K^{\text{C}}_{\omega}$ and stride length $K^{\text{S}}_{\omega}$. Zero-padding is also applied when $\lceil\frac{\lceil\sqrt{N_{\text{d}}}\rceil}{K_{\text{S}}}\rceil > \frac{\lceil\sqrt{N_{\text{d}}}\rceil}{K_{\text{S}}}$.

\subsubsection{Hidden Layers}
The second layer of the FD-CNN is also a convolutional layer but with only $2U$ convolutional kernels. This layer maps the output of the first convolutional layer,  $\tilde{\mathcal{S}}_{}^{\text{Conv1}}$, which contains $N^{\text{Conv1}}$ matrices, to an output $\tilde{\mathcal{S}}_{}^{\text{Conv2}} = [\tilde{\boldsymbol{S}}_{0}^{\text{Conv2}},\cdots,\tilde{\boldsymbol{S}}_{2U}^{\text{Conv2}}]$. This operation serves a similar purpose as the \acp{dft} in FD-GMP and FD-CNN.  Similar to~\eqref{eq:FDCNN_Conv1}, the $2u$-th output matrix $\tilde{\boldsymbol{S}}_{2u}^{\text{Conv2}}$ is computed as:
\begin{align}
    \tilde{\boldsymbol{S}}_{2u}^{\text{Conv2}} = f^{\sigma}(f^{\text{Conv2}}_{2u}(\tilde{\mathcal{S}}^{\text{Conv1}})).\label{eq:FDCNN_Conv2}
\end{align}

\subsubsection{Output Layer}
 After the second convolutional layer, each pair of matrices $\{\tilde{\boldsymbol{S}}_{2u}^{\text{Conv2}}, \tilde{\boldsymbol{S}}_{2u+1}^{\text{Conv2}}\}$ for UE $u$ is flattened and fed into a linear fully connected layer with $2N_{\text{d}}$ neurons. This layer produces the real and imaginary parts of the  predistorted symbol vector at the $u$-th UE, $\tilde{\boldsymbol{s}}_u \in \mathbb{C}^{N_{\text{d}}}$. In total, the last layer has $U$ parallel fully-connected layers, each with $2N_{\text{d}}$ neurons.

Similar to FD-NN,  FD-CNN implicitly considers the \ac{imd} products between UEs through its \ac{nn} structure to cancel the \ac{imd} between UEs. This means that it has the same adaptability problem as FD-NN when the precoder changes among coherent channels. Specifically, the first two convolutional layers of FD-CNN play a more significant role in handling the cancellation of \ac{imd} products compared to the final fully connected layer, primarily because of their nonlinear activation functions as in~\eqref{eq:FDCNN_Conv1} and \eqref{eq:FDCNN_Conv2}.

\Rev{
\vspace{-8mm}\subsection{Proposed FD-CNN: Parameter Learning} 
}
\RevMaj{Training FD-CNN and other FD-DPDs presents unique challenges not found in TD-DPDs due to fundamental differences in domain and dimensionality between the FD-DPD input and the PA output introduced by components like digital precoders and IDFTs. Specifically, the FD-DPD operates in the frequency domain and interfaces before the precoder with the UE dimensions, while the PA output is in the time domain with the antenna dimensions. These mismatches make it non-trivial to directly apply feedback schemes designed for TD-DPDs such as the \ac{ila}~\cite{eun1997new}. Additionally, obtaining the desired FD-DPD output is also challenging due to domain and dimension mismatches, making it not straightforward to use algorithms like \ac{ilc} for FD-DPDs~\cite{chani2016iterative}.}

\begin{algorithm}[t]
\caption{\Rev{Proposed FD-DPD Learning Algorithm}}\label{alg:FD_DPD_learn}
\Rev{
\KwIn{\RevMaj{$\{\check{\boldsymbol{s}}^{\text{DPD}}[k]\}$}, $\{\check{\boldsymbol{x}}[n]\}$, $\{\check{\boldsymbol{W}}[k]\}$, $\{\boldsymbol{{\theta}}^{\mathrm{GMP-PA}}_b\}$}
\KwResult{$\boldsymbol{\acute{\theta}}^{\mathrm{DPD}}$}
\While{$\mathcal{L}\big(\boldsymbol{\acute{\theta}}^{\mathrm{DPD}}\big) > \epsilon$}{
    Generate new realizations of QAM symbols \RevMaj{$\{\check{\boldsymbol{s}}^{\text{DPD}}[k]\}$}\;
    Acquire FD-DPD output \RevMaj{$\{\check{\boldsymbol{s}}^{\text{}}[k]\}$} by feeding \RevMaj{$\{\check{\boldsymbol{s}}^{\text{DPD}}[k]\}$}\;
    Apply precoding via~\eqref{eq:precod_FD}\;
    Perform IDFT of OFDM via~\eqref{eq:IDFT_precod}\;
    Obtain \RevMaj{the nonlinear PA output $\{{\boldsymbol{x}}_b^{\text{PA}}\}$} via~\eqref{eq:PA_GMP_b} using pre-trained GMP-based PAs\;
    Calculate MSE loss $\mathcal{L}(\boldsymbol{\acute{\theta}}^{\mathrm{DPD}})$ via~\eqref{eq:loss_func_MSE}\;
    Update $\boldsymbol{\acute{\theta}}^{\mathrm{DPD}}$ via backpropagation\;
}
\Return $\boldsymbol{\acute{\theta}}^{\mathrm{DPD}}$\;
}
\end{algorithm}
\RevMaj{
We propose a novel FD-DPD learning algorithm for FD-DPDs with differentiable parameters, outlined in Algorithm~\ref{alg:FD_DPD_learn}. This algorithm leverages GMP-based PA models, which can be pre-trained using the least squares method with PA output measurements through a feedback scheme. The same PA measurements are used in the ILA for TD-DPD optimization, thus both the proposed and the ILA algorithms rely equally on PA measurements, making them fair for comparison. Meanwhile, similar to the ILA for TD-DPD optimization, to handle PA variations due to environmental factors like temperature changes, periodic PA measurements are needed for the adaptation of the GMP-based PA model. This feedback-driven adaptation ensures that the pre-trained GMP-based PA models remain robust by adjusting to real-time measurements, maintaining performance against temperature drift and noise. These differentiable GMP-based PA models enable supervised learning to update the FD-DPD parameters, effectively overcome the dimension mismatch problem of FD-DPD and the PAs. Although TD-DPDs can also be optimized using pre-trained PA models, the more common method is the ILA, which, like our algorithm, also uses PA output measurements for DPD adaption. Both our learning algorithm and the ILA require adaptation to manage PA variations, making our method equally suitable for practical deployment.
}

Supervised learning is used to update the FD-DPD parameters by minimizing the \ac{mse} loss function between the PA outputs with and without any FD-DPDs:
\begin{align}
\mathcal{L}\Big(\boldsymbol{\acute{\theta}}^{\mathrm{DPD}}\Big)=\mathbb{E}_{{x}}\left\{\left|{x}_b^{\text{PA}}[n]-\acute{{x}}_b^{\text{PA}}[n]\right|^2\right\}. \label{eq:loss_func_MSE}
\end{align}
Here \RevMaj{ $\boldsymbol{\acute{\theta}}^{\mathrm{DPD}}$ represent a general form of any FD-DPD parameters, which could be the NN weights and bias for FD-CNN or FD-NN, or the GMP parameters of TD-GMP. $\acute{{x}}_b^{\text{PA}}[n]$ denotes the desired $b$-th PA output at time sample $n$, linearly amplified without nonlinearity and DPD, which is obtained by applying the desired power gain to $\{{u}_b[n]\}$ for each PA input.} Similar to the \ac{ila} and the training method in~\cite{tarver2021virtual}, this training method requires collecting each PA output $\{{{x}}_b^{\text{PA}}[n]\}$. The expectation in~\eqref{eq:loss_func_MSE} is taken over all $B$ PAs' outputs. Training uses the \ac{sgd} method with the Adam optimizer~\cite{kingma2014adam} and continues until convergence. For the isotropic scattering channel scenario,  new channel realizations and corresponding \ac{zf} precoding are generated for each training mini-batch to ensure FD-CNN is not dependent on specific channel conditions and precoding schemes. In contrast, for the \ac{los} channel scenario, the channel and precoding remain fixed during training.

\subsection{Proposed FD-CNN: Complexity Analysis and Implementation Discussion}\label{section:Complexity_Analyze}

In this section, we provide a complexity analysis and a practical implementation discussion of various DPDs.

\subsubsection{Complexity Analysis}
When dealing with DPD algorithm design, two aspects of complexity must be considered: the \textit{running complexity} and the \textit{training complexity}. We prioritize evaluating running complexity because it dominates the total computational expenses and can be quantified precisely, unlike training complexity, which is difficult to quantify accurately due to its dependence on the choice of optimization algorithms.
The number of \acp{flop} offers a precise metric, accounting for every addition, subtraction, and multiplication operation, applicable to both Volterra series-based models~\cite{moon2011enhanced} and  \ac{nn}s~\cite{liu2004dynamic, wu2021low} for DPD\footnote{The complexity of \ac{nn} activation functions like tanh, sigmoid, and ReLU varies with the hardware architecture. ReLU, selected for its minimal hardware complexity in FPGAs or ASICs, equates to one FLOP per operation.}, so it provides a fair basis for comparison across different schemes without examining specific implementation details. In this paper, we employ the same FLOP calculation method as in~\cite[Table I]{complexityGMP}.

\begin{table}
\caption{Approximate computational complexity of various DPD schemes in massive MU-MIMO-OFDM per QAM symbol in terms of the number of \acp{flop}. This MIMO system is configured with $B$ BS antennas and $U$ UEs, operating with an \ac{osr} $R$ at the BS. In addition, $V$ victim receivers are considered in FD-GMP for the generation of cancellation beams. }

    \centering
    \begin{tabular}{c|c}
    \hline
        \textbf{DPD Scheme} & \textbf{Approx. Complexity [FLOPs]} \\ \hline
       TD-GMP~\cite{GMP_2006}  & $C_{\text{Samp}}^{\text{TD-GMP}}
       RB$\\
       FD-GMP~\cite{brihuega2021frequency, brihuega2022beam}  & $(C_{\text{Samp}}^{\text{TD-GMP}}+8B)R(U+V)$\\
       FD-NN~\cite{tarver2021virtual}  & $(D+D^2/U+8B)RU$ \\
       FD-CNN  & $8(K_{\text{C}}^2 N^{\text{Conv1}} + N^{\text{d}})U$ \\
    \end{tabular}
    \label{tab:FLOPs_approx}
\end{table}

A summary of the approximated complexity of DPD schemes, TD-GMP, FD-GMP, and FD-NN in digital massive MU-MIMO-OFDM are shown in
Table~\ref{tab:FLOPs_approx}. The detailed complexity calculations of these schemes are given in Appendix~\ref{appdx:complexity}, respectively.  
Based on the approximation summarized in Table~\ref{tab:FLOPs_approx}, the benefits and drawbacks of the different DPD schemes are discussed and compared in the following. 
\begin{itemize}
    \item TD-GMP: The number of FLOPs grows linearly with the number of \ac{bs} antennas $B$ and the \ac{osr} $R$. Such linear escalation presents a complexity challenge in massive MU-MIMO systems, especially when dealing with a large number of antennas and a wide bandwidth. 
    \item FD-GMP: Operating in the FD before the precoder, FD-GMP's complexity scales linearly with the number of UEs plus \ac{imd} beams for victim receivers, i.e., ($U+V$), instead of the number of antennas $B$. Yet, the additional precoding of OOB subcarriers, after the DFTs, incurs extra costs, making the required \acp{flop} grow with $B$. This rise is exacerbated with more antennas. Hence, FD-GMP does not fully address the complexity problem of TD-GMP. 
    \item FD-NN: Operating before the precoder, FD-NN's complexity scales linearly with the number of UEs $U$ and the OSR $R$. Compared with FD-GMP, FD-NN saves the computational cost of \ac{imd} beams as they are implicitly considered in the \ac{nn} structure. Similar to FD-GMP, FD-NN also requires additional precoding costs due to (I)DFTs usage, eventually leading to a linear increase in computational complexity with the number of BS antennas $B$, mirroring the complexity issues faced by FD-GMP and TD-GMP in large antenna arrays.   
\Rev{
\item FD-CNN: Unlike TD-based DPD, FD-CNN operates in the FD before the precoder, making its computational complexity scale with the number of UEs $U$ rather than the number of BS antennas $B$. Compared to FD-GMP and FD-NN, FD-CNN utilizes low-complexity convolutional filters instead of (I)DFTs, allowing for flexible input/output size adjustments and non-oversampled processing. This approach eliminates the need for additional precoding on oversampled subcarriers and thus detaches its complexity from the number of antennas $B$, which makes FD-CNN advantageous for large arrays. However, the complexity of FD-CNN increases with the number of data subcarriers $N_{\text{d}}$ due to its fully-connected layer, which can be challenging in wide bandwidth scenarios.
}

\end{itemize}

\RevMaj{
\subsubsection{Practical Implementation}
DPD implementations are commonly done using \acp{fpga} or \acp{asic}~\cite{gilabert2008multi,ma2013fpga,tarver2019FPGAdesign,li2024gpu_GMP}. \Acp{fpga} are preferred for prototyping due to their flexibility and reconfigurability, while ASICs are more power-efficient for large-scale deployment but require specific hardware design, limiting their adaptability. As signal bandwidths increase and higher OSRs are needed to prevent aliasing, throughput becomes a critical hardware bottleneck. This is especially true for TD-DPDs, like TD-GMP, whose complexity grows with the number of antennas. FD-DPDs, such as the proposed FD-CNN DPD, offer reduced computational complexity and lower power consumption in large antenna systems, though they still face challenges with high-bandwidth signals due to increased input size.}

\RevMaj{A recent work~\cite{li2024gpu_NN} has shown that \acp{gpu} can meet the high throughput requirements for a wideband signal and an NN-based DPD with a large number of parameters, although GPUs generally are less power efficient than FPGAs and ASICs. Moreover, for training DPD models, especially NN-based DPDs that require longer training times, GPUs are much more efficient than FPGAs or ASICs, which are less suitable for training tasks due to their fixed nature. In the future, DPD implementations are likely to use a combination of ASICs, FPGAs, and GPUs: ASICs for energy-efficient large-scale deployment, FPGAs for flexible prototyping and adaptable designs, and GPUs for both training and high-throughput inference in wideband environments.
}

\Rev{
\vspace{-1mm}\subsection{Qualitative Comparison of DPD Strategies}
We next analyze the benefits and drawbacks of various DPD methods, focusing on their methodology, deployment, scalability to BS antennas and UEs, parameter learning methods, adaptability to different channel types, and \ac{imd} handling. Key findings are summarized in Table~\ref{tab:DPD_compare_aspects}.  The most important aspects-- i.e., scalability to BS antennas and UEs, parameter learning methods, adaptability to different channel types, and \ac{imd} handling -- are discussed in the following three subsubsections.
}
\Rev{
\subsubsection{Scalability with Antennas and UEs}
With an increasing number of antennas, TD-GMP scales poorly due to its per RF chain deployment. This limitation persists even with complexity-reducing implementations such as shared-DPD~\cite{shared_dpd} and low-sampling rate DPD~\cite{li2019lowsampling}, which depend on analog or hybrid precoding. FD-GMP and FD-NN offer better scalability by being deployed before digital precoding. However, the additional precoding costs associated with using (I)DFTs eventually limit their scalability. The proposed FD-CNN addresses this issue by replacing (I)DFTs with CNNs, thereby eliminating the additional precoding costs and achieving superior scalability with an increasing number of antennas. With an increasing number of UEs, TD-GMP’s scalability is unaffected, while FD-GMP, FD-NN, and FD-CNN exhibit poor scalability due to linear scaling with UEs. However, in massive MIMO, where UEs are fewer than antennas, the impact on FD-DPD scalability is moderate.
}
\Rev{
\subsubsection{Parameter Learning}
Obtaining the desired DPD output is challenging because the desired PA input cannot be straightforwardly determined. For TD-GMP, this issue is addressed using the \ac{ila}\cite{eun1997new}, which requires dedicated feedback paths by treating the PA output as the input and the PA input as the desired output. FD-DPDs (FD-GMP, FD-NN, and FD-CNN) face greater difficulties due to domain and dimension changes caused by digital precoders and IDFTs. The \ac{ota} learning method, via an observation receiver as in\cite{brihuega2022beam}, can be used to acquire \ac{ota} signals, but it requires knowledge of the channels. Another method involves converting TD-DPD output back to the FD, as in~\cite{tarver2021virtual}, which also relies on channel knowledge and is limited by the TD-DPD model’s accuracy. Alternatively, the proposed learning methodology in Algorithm~\ref{alg:FD_DPD_learn} can be applied to any FD-DPDs with differentiable parameters.
}

\Rev{
\subsubsection{\ac{imd} Handling in Different Channels with Multiple UEs}
In frequency-flat LOS channels, TD-GMP effectively addresses distortion for each PA, demonstrating high capability in handling \acp{imd} for both single or multiple UEs. The three FD-DPDs (FD-GMP, FD-NN, and FD-CNN) also handle IMDs effectively in frequency-flat LOS channels, though their capability degrades with more UEs. Specifically, the effectiveness of FD-DPDs depends highly on the PA radiation pattern. Unlike FD-GMP, which explicitly generates IMD products, FD-NN and FD-CNN implicitly handle \ac{imd} products within their \ac{nn} layers, thereby avoiding additional IMD products at the output. 
}
\Rev{
For isotropic scattering channels, TD-GMP adjusts its operation power based on input levels to maintain the high capability of IMD handling. Although the three FD-DPDs (FD-GMP, FD-NN, FD-CNN) cannot address each IMD beam individually due to their large number, the relaxation of the \ac{oob} linearization requirement allows FD-DPDs to be valuable, provided that they satisfy the in-band linearization requirement at the UEs. 
}

\begin{table*}[th]
\centering
\caption{\Rev{Qualitative comparison of state-of-the-art TD and FD DPDs with proposed FD-CNN. The quantification of the IMD cancellation capability of different DPDs, together with their linearization capability under crosstalk, are analyzed in Section V-C.}} \label{tab:DPD_compare_aspects}
\Rev{
\begin{tabularx}{0.92\textwidth}{Xccccc}
\cline{1-6}
\multicolumn{2}{c}{}                                                                                             & TD-GMP~\cite{GMP_2006}                                                     & FD-GMP~\cite{brihuega2022beam}                                                                 & FD-NN~\cite{tarver2021virtual}                                                                  & FD-CNN (This work)                                                     \\ \cline{1-6}
\multicolumn{2}{l|}{Methodology}                                                                                                         & GMP in TD                                                  & \begin{tabular}[c]{@{}c@{}}GMP in FD\\ via (I)DFTs\end{tabular}        & \begin{tabular}[c]{@{}c@{}}NN in FD \\ via (I)DFTs\end{tabular}        & \begin{tabular}[c]{@{}c@{}} \ac{nn} in FD\\ via CNNs\end{tabular}      \\ \cline{1-6}
\multicolumn{2}{l|}{DPD Dimension} 
                                                                                              & \begin{tabular}[c]{@{}c@{}}One per antenna\end{tabular} & \begin{tabular}[c]{@{}c@{}}One per UE\end{tabular}             & \begin{tabular}[c]{@{}c@{}}One fo all UEs\end{tabular}        & \begin{tabular}[c]{@{}c@{}}One for all UEs\end{tabular}     \\ \cline{1-6}
\multicolumn{2}{l|}{Additional Precoding Cost}                                                                                     & No                                                         & \begin{tabular}[c]{@{}c@{}}Yes: \\ due to (I)DFTs\end{tabular}         & \begin{tabular}[c]{@{}c@{}}Yes: \\ due to (I)DFTs\end{tabular}         & \begin{tabular}[c]{@{}c@{}}No: \\ due to CNNs\end{tabular}             \\ \cline{1-6}
\multicolumn{2}{l|}{Scalability with Antennas}                                                                                  & Poor                                                       & Moderate                                                               & Moderate                                                               & High                                                                   \\ \cline{1-6}
\multicolumn{2}{l|}{Scalability with UEs}                                                                                  & High                                                       & Moderate                                                               & Moderate                                                               & Moderate                                                                   \\ \cline{1-6}
\multicolumn{2}{l|}{Parameter Learning Algorithm}                                                                                                & ILA~\cite{eun1997new}                                                        & \begin{tabular}[c]{@{}c@{}}1. OTA~\cite{brihuega2022beam} 2.~\cite{tarver2021virtual}\\ 3. Algorithm~\ref{alg:FD_DPD_learn}\end{tabular} & \begin{tabular}[c]{@{}c@{}}1. OTA~\cite{brihuega2022beam} 2.~\cite{tarver2021virtual} \\ 3.  Algorithm~\ref{alg:FD_DPD_learn}\end{tabular} & \begin{tabular}[c]{@{}c@{}}1. OTA~\cite{brihuega2022beam} 2.~\cite{tarver2021virtual}\\ 3.~Algorithm 1\end{tabular} \\ \cline{1-6}
\multicolumn{1}{c|}{\multirow{4}{*}{ \begin{tabular}[c]{@{}c@{}}IMD \\ Cancellation\end{tabular} }} & \multicolumn{1}{c|}{LOS $\&$ Single UE}
       & \begin{tabular}[c]{@{}c@{}}High\end{tabular}        & \begin{tabular}[c]{@{}c@{}}High\end{tabular}                & \begin{tabular}[c]{@{}c@{}}High\end{tabular}                & \begin{tabular}[c]{@{}c@{}}High\end{tabular}                \\ \cline{2-6} 
\multicolumn{1}{c|}{} 
&\multicolumn{1}{c|}{LOS $\&$ Multiple UEs}                 & \begin{tabular}[c]{@{}c@{}}High\end{tabular}        & \begin{tabular}[c]{@{}c@{}}Moderate\end{tabular}                & \begin{tabular}[c]{@{}c@{}}Moderate\end{tabular}                & \begin{tabular}[c]{@{}c@{}}Moderate\end{tabular}                \\ \cline{2-6} 
\multicolumn{1}{c|}{} 
&\multicolumn{1}{c|}{Isotropic Scattering $\&$ Single UE} & \begin{tabular}[c]{@{}c@{}}High\end{tabular}        & \begin{tabular}[c]{@{}c@{}}Moderate\end{tabular}                & \begin{tabular}[c]{@{}c@{}}Moderate\end{tabular}                & \begin{tabular}[c]{@{}c@{}}Moderate\end{tabular}                 \\ \cline{2-6} 
\multicolumn{1}{c|}{} 
&\multicolumn{1}{c|}{Isotropic Scattering $\&$ Multiple UEs}& \begin{tabular}[c]{@{}c@{}} High\end{tabular}        & \begin{tabular}[c]{@{}c@{}}Poor\end{tabular}                & \begin{tabular}[c]{@{}c@{}} Poor\end{tabular}                & \begin{tabular}[c]{@{}c@{}}Poor\end{tabular}                 \\ \cline{1-6}
\end{tabularx}
}
\vspace{-0.2cm}
\end{table*}

\section{Numerical Results}
In this section, we will present the quantitative analysis in terms of the complexity and performance of the different DPD approaches. We first detail the DPD figures of merit, then describe the simulation setup, and finally provide our results with a discussion. 

\vspace{-4mm}\subsection{Figures of Merit for Linearity in Massive MIMO}
We aim to ensure sufficient in-band power to attain the desired SINR at the UEs while minimizing out-of-band radiation to avoid interference with other systems. In massive MIMO systems, conventional \ac{oob} metric can be too stringent, because the in-band signal can obtain higher power gain at the UEs than out-of-band distortion, thanks to beamforming. In this section, we adopt the \ac{evm} as a metric to assess the in-band signal quality and several versions of \ac{aclr} as metrics to evaluate out-of-band linearization.
 
\subsubsection{In-band Quality by EVM}
The in-band signal quality is commonly quantified by evaluating the \ac{evm} of the received  signal as in~\cite[Eq. (2.4)]{EVM_BER}.

\subsubsection{Out-of-band Distortion Measurement by TRP-ACLR}
The conventional \ac{aclr}, measured at the Tx, is often too stringent for large arrays~\cite{mollen2018spatial}. We adopt an alternative scalar-valued \ac{trp}-\ac{aclr}, which is defined in 3GPP~\cite{3gpp-TS-38.104}. It is the ratio between the \ac{trp} of the signal in the main channel to the \ac{trp} of the signal in the adjacent channel, summed over all directions.

\vspace{-4mm}\subsection{Simulation Setup}
\subsubsection{Parameters}
We consider a set of simulated PAs using the GMP model in~\eqref{eq:PA_GMP_b} with nonlinear order $Q=7$, memory length $M=5$, and cross-term length $G=1$. These parameters are estimated using real measurements from the RF WebLab using a $100$ MHz OFDM signal~\cite{landin2015weblab} with a $200$ MHz sampling rate. All PAs are assumed to have the same parameters. The GMP-based PA coefficients are fixed over time. The saturation point and measurement noise standard deviation of each PA are $24.02$ V $(\approx 37.6$ dBm with a $50$ $\Omega$ load impedance) and $0.053$ V, respectively. The digital precoder adopts the \ac{zf} precoding~\eqref{eq:Zero_forcing}. We consider a $30.7$ MHz OFDM setup with a subcarrier spacing $\Delta_f=120$ kHz, number of data subcarriers $N_{\text{d}}=256$, \ac{osr} $R=4$ by an IDFT size $N=1024$.
\subsubsection{Channel scenarios}
Two different channel scenarios are considered based on the carrier frequency: frequency-selective isotropic scattering and frequency-flat \ac{los}. 


We model an isotropic fading channel using the uncorrelated Rayleigh fading channel model~\eqref{eq:channel_Rayleigh} with $\gamma=100$ channel taps. The large-scale fading coefficient $\sigma_{\beta_{\boldsymbol{p}}}^2$ is modeled in decibels and set to the same value as~\cite{bjornson2017massive}
\begin{align}
    \sigma_{\beta_{\boldsymbol{p}}}^2 [\text{dB}]=\Upsilon-  10\alpha \log _{10}\left({\| \boldsymbol{p} - \boldsymbol{p}_{\text{BS}}\|}/{1 \mathrm{~m}}\right)+F_{p},
\end{align}
where the pathloss exponent is $\alpha=3.76$, the median channel gain at a reference distance of $1$ m is $\Upsilon=-35.3$dB, and the shadow fading is modeled by $F_{p}  \sim  \mathcal{N}(0,\sigma_{\text{sf}}^2)$  with the standard deviation $\sigma_{\text{sf}}=4$. These propagation parameters match well with the NLoS macro cell 3GPP model for $2$ GHz carriers~\cite[Table B.1.2.1-1]{3gpp-TS-36.814}. All users are randomly placed around the array with the same distance to the BS of $25$ m. The thermal noise variance is $-174$ dBm/Hz, which yields a total receiver noise power of $\sigma^2=-92.1$ dBm including a receiver noise figure $7$ dB. An average transmit power of the BS is $P_{\text{BS}}=48$ dBm, which gives an average output power of each PA $\sum_{b=0}P_b^{\text{PA}}/B=28$ dBm. 

We model a frequency-flat \ac{los} channel using~\eqref{eq:channel_LOS}, setting the carrier frequency $f_c = 30$ GHz. The large-scale fading coefficient $\beta_{\boldsymbol{p}}^2$ is set according to the LOS urban microcell street Canyon 3GPP  model in~\cite[Table 7.4.1-1]{3gpp_TS_38_901} with median channel gain $\Upsilon=-61.9$ dB at $1$m, pathloss $\alpha=2.1$, and shadowing fading standard deviation $\sigma_{\text{sf}}=4$. The same noise power is used as in the isotropic fading scenario.

\subsubsection{DPD Coefficients Identification}
For TD-GMP, the GMP coefficients for DPD at each RF antenna are estimated separately using the \ac{ila}~\cite{eun1997new} via the least squares algorithm. \Rev{For a fair comparison, all FD-based DPD schemes, including the proposed FD-CNN, utilize the same training approach proposed in Algorithm~\ref{alg:FD_DPD_learn} and the Adam optimizer with a learning rate $0.001$. The number of OFDM symbols for each mini-batch and the maximum number of batches are $100$ and $5000$. While fine-tuning the neural network and training process for FD-CNN could yield even better performance, this is left for future work. For FD-NN, we follow the setup in~\cite{tarver2021virtual} with 1 hidden layer and 15 neurons. For the proposed FD-CNN, the number of convolution kernels is $N^{\text{Conv1}}=20$, the kernel size is $K_{\text{C}}=3$, and the stride side is $1$.} 

\vspace{-4mm}\subsection{Simulation results}

\subsubsection{Complexity Versus Number of BS Antennas}\label{section:results_complexity_vs_B_U}
\begin{figure}[t]
\centering
    \begin{subfigure}{0.5\textwidth}
\begin{tikzpicture}[font=\scriptsize]
\definecolor{color0}{rgb}{0.12156862745098,0.466666666666667,0.705882352941177}
\definecolor{color1}{rgb}{1,0.498039215686275,0.0549019607843137}
\definecolor{color2}{rgb}{0.172549019607843,0.627450980392157,0.172549019607843}
\definecolor{color4}{rgb}{0.83921568627451,0.152941176470588,0.156862745098039}
\definecolor{color3}{rgb}{0.580392156862745,0.403921568627451,0.741176470588235}
\definecolor{color5}{rgb}{0.549019607843137,0.337254901960784,0.294117647058824}
\definecolor{color6}{rgb}{0.890196078431372,0.466666666666667,0.76078431372549}
\definecolor{color7}{rgb}{0.737254901960784,0.741176470588235,0.133333333333333}

\begin{axis}[
width=8.5cm,
height=5.cm,
legend cell align={left},
legend style={
  fill opacity=1,
  draw opacity=1,
  text opacity=1,
  at={(0,1)},
  anchor=north west
},
log basis x={2},
log basis y={10},
tick align=outside,
tick pos=left,  
x grid style={white!69.0196078431373!black},
xlabel={Number of antennas, $B$},
xmajorgrids,
xmin=1, xmax=4096,
xmode=log,
xtick={1,4,16,64,256,1024,4096},
xticklabels={1,4,16,64,256,1024,4096},
xtick style={color=black},
y grid style={white!69.0196078431373!black},
ylabel={Total number of FLOPs},
ymajorgrids,
yminorgrids,
ymin=1e3, ymax=2e7,
ymode=log,
ytick style={color=black},
ytick={1e3,1e4,1e5,1e6,1e7}
]
\addplot [semithick, color0, line width=1.0pt, mark=o, mark size=2, mark options={solid,fill opacity=0}]
table {%
1 994
4 3976
16 15904
100 99400
256 254464
1024 1017856
4096 4071424
};
\addlegendentry{$C^{\text{TD-GMP}}$~\cite{GMP_2006} ($R=1$)}
\addplot [semithick, color1, line width=1.0pt, mark=o, mark size=2, mark options={solid,fill opacity=0}]
table {%
1 3976
4 15904
16 63616
100 397600
256 1017856
1024 4071424
4096 16285696
};
\addlegendentry{$C^{\text{TD-GMP}}$~\cite{GMP_2006} ($R=4$)}
\addplot [semithick, color2, line width=1.0pt, mark=o, mark size=2, mark options={solid,fill opacity=0}]
table {%
1 4080
4 4152
16 4440
100 6456
256 10200
1024 28632
4096 102360
};
\addlegendentry{$C^{\text{FD-GMP}}$~\cite{brihuega2022beam}}
\addplot [semithick, color3, mark=square, mark size=2, line width=1.0pt, mark options={solid,fill opacity=0}]
table {%
1 4904
4 4976
16 5264
100 7280
256 11024
1024 29456
4096 103184
};
\addlegendentry{$C^{\text{FD-NN}}$~\cite{tarver2021virtual}}
\addplot [semithick, color4, line width=1.0pt, mark=triangle, mark size=2.5, mark options={solid,fill opacity=0}]
table {%
1 1800
4 1800
16 1800
100 1800
256 1800
1024 1800
4096 1800
};
\addlegendentry{$C^{\text{FD-CNN}}$}
\end{axis}

\end{tikzpicture}
     \vspace{-0.cm}
\caption{FLOPs versus the number of BS antennas. $U=1$. }
	\label{fig:FLOPs_vs_B}
    \end{subfigure}
\hfill
     \vspace{-0.cm}
	 \begin{subfigure}{0.5\textwidth}
\begin{tikzpicture}[font=\scriptsize]
\definecolor{color0}{rgb}{0.12156862745098,0.466666666666667,0.705882352941177}
\definecolor{color1}{rgb}{1,0.498039215686275,0.0549019607843137}
\definecolor{color2}{rgb}{0.172549019607843,0.627450980392157,0.172549019607843}
\definecolor{color4}{rgb}{0.83921568627451,0.152941176470588,0.156862745098039}
\definecolor{color3}{rgb}{0.580392156862745,0.403921568627451,0.741176470588235}
\definecolor{color5}{rgb}{0.549019607843137,0.337254901960784,0.294117647058824}
\definecolor{color6}{rgb}{0.890196078431372,0.466666666666667,0.76078431372549}
\definecolor{color7}{rgb}{0.737254901960784,0.741176470588235,0.133333333333333}

\begin{axis}[
width=8.5cm,
height=5.cm,
legend cell align={left},
legend style={
  fill opacity=1,
  draw opacity=1,
  text opacity=1,
  at={(1,0)},
  anchor=south east
},
log basis x={2},
log basis y={10},
tick align=outside,
tick pos=left,  
x grid style={white!69.0196078431373!black},
xlabel={Number of UEs, $U$},
xmajorgrids,
xmin=1, xmax=64,
xmode=log,
xtick={1,2,4,8,16,64},
xticklabels={1,2,4,8,16,64},
xtick style={color=black},
y grid style={white!69.0196078431373!black},
ylabel={Total number of FLOPs},
ymajorgrids,
yminorgrids,
ymin=1e3, 
ymax=6e5,
ymode=log,
ytick style={color=black},
]
\addplot [semithick, color0, line width=1.0pt, mark=o, mark size=2, mark options={solid,fill opacity=0}]
table {%
1 99400
2 99400
4 99400
8 99400
16 99400
64 99400
};
\addlegendentry{$C^{\text{TD-GMP}}$~\cite{GMP_2006} ($R=1$)}
\addplot [semithick, color1, line width=1.0pt, mark=o, mark size=2, mark options={solid,fill opacity=0}]
table {%
1 397600
2 397600
4 397600
8 397600
16 397600
64 397600
};
\addlegendentry{$C^{\text{TD-GMP}}$~\cite{GMP_2006} ($R=4$)}
\addplot [semithick, color2, line width=1.0pt, mark=o, mark size=2, mark options={solid,fill opacity=0}]
table {%
1 6456
2 12912
4 25824
8 51648
16 103296
64 413184
};
\addlegendentry{$C^{\text{FD-GMP}}$~\cite{brihuega2022beam}}
\addplot [semithick, color3, mark=square, mark size=2, line width=1.0pt, mark options={solid,fill opacity=0}]
table {%
1 7280
2 14560
4 29120
8 58240
16 116480
64 465920
};
\addlegendentry{$C^{\text{FD-NN}}$~\cite{tarver2021virtual}}
\addplot [semithick, color4, line width=1.0pt, mark=triangle, mark size=2.5, mark options={solid,fill opacity=0}]
table {%
1 1800
2 3600
4 7200
8 14400
16 28800
64 115200
};
\addlegendentry{$C^{\text{FD-CNN}}$}
\end{axis}

\end{tikzpicture}
	      \vspace{-0.2cm}
	\caption{FLOPs versus the number of UEs. $B=100$. }
	\label{fig:FLOPs_vs_U}
    \end{subfigure}
	\vspace*{-0.1cm}
	\caption{The number of FLOPs required for each DPD scheme of each QAM symbol versus the number of BS antennas $B$ and UEs $U$. }
\end{figure}
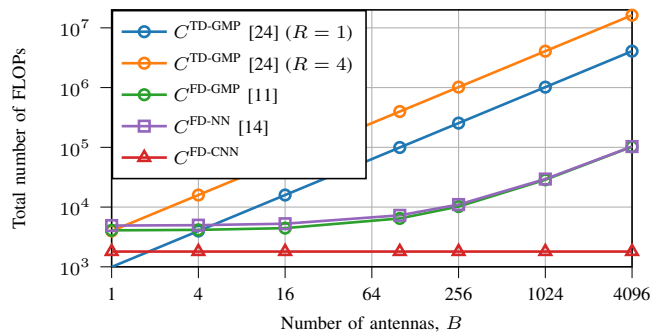
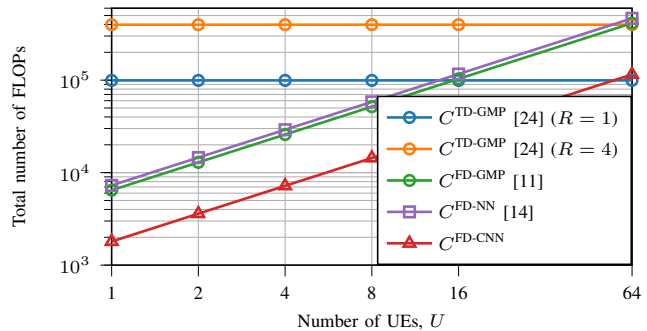

Fig.~\ref{fig:FLOPs_vs_B} shows the number of \acp{flop} required per QAM symbol for a single UE as a function of the number of BS antennas $B\in \{1,10,100,1000\}$. The considered DPD schemes with an OSR $R=4$ are TD-GMP, FD-GMP, and FD-NN, while the considered DPD schemes with an OSR $R=1$ are  TD-GMP and the proposed FD-CNN. Note that the introduced additional precoding cost associated with OOB subcarriers are included for FD-GMP and FD-NN.

As the number of BS antennas increases, the number of \acp{flop} for TD-GMPs, $C^{\text{TD-GMP}}$, with OSR $R=\{1,4\}$ grows linearly as each RF chain is associated with a dedicated TD-GMP. In contrast, all three FD-DPD schemes, i.e., FD-GMP, FD-NN, and FD-CNN, alleviate this computational complexity problem to varying degrees. Specifically, the number of \acp{flop} for FD-GMP, $C^{\text{FD-GMP}}$, and FD-NN, $C^{\text{FD-NN}}$, eventually increases linearly with $B$ as the additional cost of OOB precoding starts to dominate, limiting the saved  number of \acp{flop}. However, the number of FLOPs for the proposed FD-CNN, $C^{\text{FD-CNN}}$, remains completely invariant to $B$, making it an attractive option for very large antenna systems. Compared with FD-NN and TD-GMP, FD-CNN saves around $2.2\times$ and $8.9\times$ FLOPs for $B = 100$ and $U = 1$, respectively, and these savings can be further increased up to $26\times$ and $286\times$ when $B$ increases to $1000$, respectively. \RevMaj{However, it is important to note that, with current technology, using more than 100 antennas in a fully digital array is impractical due to not only the high computational complexity of DPD but also the power demands of other RF chain components such as DACs. In such cases, hybrid or analog MIMO systems are more feasible.}

Moving on to the comparison of the computational complexity in a MU MIMO system, Fig.~\ref{fig:FLOPs_vs_U} shows the number of \acp{flop} required per QAM symbol as a function of $U\in\{1,2,4,8,16,32,64\}$ while keeping $B$ fixed at $100$. Although TD-GMP DPDs ($R\in\{1,4\}$) require more \acp{flop} than FD-DPD schemes, their complexity remains constant with respect to $U$ since their complexities are determined by the number of antennas. In contrast, FD-based DPDs exhibit a linear increase in the number of \acp{flop} with $U$, eventually making them less computationally advantageous for large $U$ compared to TD-DPDs ($U>16$ for FD-GMP and FD-NN, and $U>32$ for FD-CNN). In such scenarios, it is more advisable to consider TD-based DPD schemes.

\begin{figure}[t]
	\centering
		\vspace*{0. \baselineskip}
	 \input{figures/OTA_PSD_B100U1}
			\vspace*{-0.5cm}
	\caption{Normalized \ac{psd} and \ac{psd} of error at an intended UE and victim w/o DPD and with different DPD schemes for $B=100$, $U=1$ in a frequency-selective isotropic scattering channel.}
	\label{fig:OTA_PSD_B100U1}
	\vspace*{-0.2cm}
\end{figure}

\begin{figure}[t]
	\centering
		\vspace*{0. \baselineskip}
	 \input{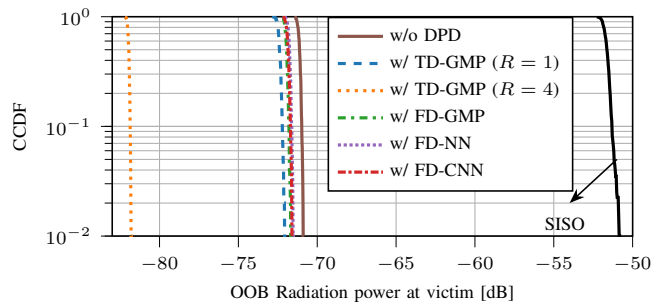}
	\caption{The distribution of the power received by a victim in the adjacent band in a frequency-selective isotropic scattering channel w/o DPD and with different DPD schemes.  The results from a \ac{siso} system  without DPD are also shown, whose transmitter power is scaled to reach the same in-band power at the UE as the MIMO system.}
	\label{fig:OOB_Distri_B100U1}
 \end{figure}
\begin{table}[t]
    \centering
    \caption{The EVM and TRP-ACLR results of different DPD schemes for a frequency-selective isotropic scattering channel with $B=100$ and $U=[1,2,4]$ in Fig.~\ref{fig:OTA_PSD_B100U1} and~\ref{fig:OOB_Distri_B100U1}.}
    \begin{tabular}{lcc}
\hline Scheme & $\text{ACLR}_{\text{TRP}}$ (dBc) & EVM (\%) \\
\hline 3GPP Requirement~\cite{3gpp-TS-38.104} & 26 & 3.5
\\ 
 w/o DPD & $[31.7, 28.8, 25.6]$ & $[6.9, 7.0, 7.1]$ \\
TD GMP, $R=4$~\cite{GMP_2006} & $[42.3, 39.3, 36.2]$ & $[0.6, 0.6, 0.7]$ \\
TD GMP, $R=1$~\cite{GMP_2006} & $[32.9, 29.8, 26.8]$ & $[1.7, 1.8,1.9]$ \\
FD GMP~\cite{brihuega2022beam}  & $[32.6, 29.6, 26.5]$ & $[1.8, 1.8,1.9]$ \\
FD  \ac{nn}~\cite{tarver2021virtual}  & $[32.5, 29.5, 26.5]$ & $[2.0, 2.1,2.2]$ \\
FD CNN  & $[32.8, 29.8, 26.7]$ & $[2.2, 2.2,2.3]$ \\
\hline
\end{tabular}
    \label{tab:EVM_ACLR_B100U1}
\end{table}

In summary, Fig.~\ref{fig:FLOPs_vs_B} and~\ref{fig:FLOPs_vs_U} provide valuable guidance for selecting an appropriate DPD scheme based on computational complexity budget for scenarios with different $B$ and $U$.

\subsubsection{Isotropic scattering scenario}

\begin{figure*}[t!]
\centering
\begin{minipage}[h]{0.48\linewidth}
\renewcommand{\figurename}{(a)}
\renewcommand{\thefigure}{}
\addtocounter{figure}{-1}
\hspace{-0.9cm}
 {\input{figures/BP_B100U1}}
\captionof{figure}[]{$U=1$ UE at distance $28$m and direction $30^{\circ}$ to the BS.}
\label{fig:BPs_U1}
\end{minipage}
\begin{minipage}[h]{0.48\linewidth}
\renewcommand{\figurename}{(b)}
\renewcommand{\thefigure}{}
\addtocounter{figure}{-1}
\centering
 \centerline{\input{figures/BP_B100U4}}
\captionof{figure}{$U=4$ UEs with distances $=\{250, 85, 48, 28\}$ m and directions $=\{-28^{\circ}, -57^{\circ}, -14^{\circ},-10^{\circ}\}$ to the BS, respectively. Pathloss-based power allocation is utilized so all UEs receive the same in-band power. UE at $-28^{\circ}$ is allocated with the dominant power. }
\label{fig:BPs_U4}
\end{minipage}
\begin{minipage}[h]{0.48\linewidth}
\renewcommand{\figurename}{(c)}
\renewcommand{\thefigure}{}
\addtocounter{figure}{-1}
\centering
{\input{figures/BP_B100U2}}
\captionof{figure}{$U=2$ UEs at distances $=\{28, 28\}$ m and directions $=\{30^{\circ}, -45^{\circ}\}$ with equal power allocation.}
\label{fig:BPs_U2}
\end{minipage}
\begin{minipage}[h]{0.48\linewidth}
\renewcommand{\figurename}{(d)}
\renewcommand{\thefigure}{}
\addtocounter{figure}{-1}
\centering
 \centerline{\input{figures/BP_B100U10}}
\captionof{figure}{$U=10$ UEs at the same distance $28$m and different directions to the BS with equal power allocation.}
\label{fig:BPs_U10}
\end{minipage}
    \caption{The beampatterns from a large linear array consist of $B=100$ antennas spaced by a half wavelength that serves $U=1,2,4,10$ UEs in a LOS channel w/o DPD and with different DPD schemes. The radiated in-band and OOB power from a SISO system $(B=1)$ with an ideal linear-clipping PA are also shown with pink lines, whose transmitter power is scaled to reach the same in-band power at the UE as the MIMO system for each subfigure.}
    \label{fig:BPs}
    \vspace{-0.3cm}
\end{figure*}
An isotropic scattering scenario modeled by uncorrelated Rayleigh fading between $B=100$ BS antennas and $U=\{1,2,4\}$ UEs, is analyzed without crosstalk. For $U=1$, the observed OTA \acp{psd} and the PSD of errors at the UE  are shown in Fig.~\ref{fig:OTA_PSD_B100U1} for the same  DPD schemes in Section~\ref{section:results_complexity_vs_B_U}. It shows that the TD-GMP DPD ($R=4$) achieves the best linearization performance for both in-band and \ac{oob} at the expense of the highest computational complexity, while all the other complexity-reduced DPD schemes achieve relatively good in-band linearization performance, i.e., with an EVM less than the 3GPP requirement $3.5\%$ for 256-QAM, as shown in Table~\ref{tab:EVM_ACLR_B100U1}.

To further assess the OOB linearization performance, Fig.~\ref{fig:OOB_Distri_B100U1} shows the distribution of power in the adjacent band received by a randomly located victim for $U=1$. The victim is randomly distributed around the BS with the same distance of $25$ m as the served UE. The OOB result from a SISO system, i.e., $B=1$, is included, with normalized transmit power to match the same in-band power as the MIMO system. We first notice that the OOB radiations for the MIMO system are nearly isotropic and significantly lower compared to that from a SISO system. The most unfortunate victim in a MIMO system ($B=100$) receives $\approx 20$dB less OOB radiation than in a SISO system ($B=1$). This reduction in OOB radiation is due to the array gain in the MIMO system and explains why conventional ACLR requirements can be relaxed in MIMO scenarios. Note that this relaxation grows linearly with the array gain. Thus, although the FD-DPD schemes exhibit slightly worse OOB linearization performance than the TD-GMP ($R=4$) as shown in Fig.~\ref{fig:OOB_Distri_B100U1}, the remaining OOB radiation at a random victim  remains acceptable, all with better TRP-ACLRs than the 3GPP requirement as shown in Table.~\ref{tab:EVM_ACLR_B100U1}. As the number of UEs equidistant from the base station increases, sharing the array gain, all schemes experience degraded TRP-ACLRs, with $3$ dB and $6$ dB for $U=2$ and $U=4$, respectively. EVM degradation is less pronounced than TRP-ACLR. With more UEs, all FD-DPDs are likely to fail in meeting the ACLR requirement.

\begin{table}[h]
    \centering
    \caption{The average EVM and TRP-ACLR results of different DPD schemes for a LOS channel with $B=100$ and $U=[1,4,2,10]$ for Fig.~\ref{fig:BPs} (a-d), respectively. Equal power allocation is used except for the case $U=4$. }
    \begin{tabular}{lcc}
\hline DPD Scheme & $\text{ACLR}_{\text{TRP}}$  (dBc) & EVM (\%) \\
\hline 3GPP Req.~\cite{3gpp-TS-38.104} & $26$ & $3.5$
\\ 
 w/o DPD & $[26.6, 25.4, 23.2, 16.2]$ & $[11.2, 15.5, 9.9, 7.9]$ \\
TD-GMP, $R=4$~\cite{GMP_2006} & $[44.5, 41.3, 40.5, 32.9]$ & $[1.3, 1.8, 1.1, 1.1]$ \\
TD-GMP, $R=1$~\cite{GMP_2006} & $[29.7, 28.2, 23.5,16.7] $ & $[6.3, 9.6, 5.3,3.4] $ \\
FD-GMP~\cite{brihuega2022beam}  & $[33.7, 33.6, 28.6,27.0]$ & $[1.8,3.3,2.6,3.2]$ \\
FD-NN~\cite{tarver2021virtual}  & $[35.2, 34.7, 34.0,27.9]$ & $[1.9,3.3,2.7,3.4]$ \\
FD-CNN  & $[35.4, 36.4, 34.9,27.5]$ & $[2.0,3.4,2.8,3.4]$ \\
\hline
\end{tabular}
    \label{tab:EVM_ACLR_LOS}
\end{table}
\vspace{-0.cm}

\begin{table}[h]
    \centering
    \caption{\Rev{The average SLL relative to the main lobe for DPDs in a LOS channel with $B=100$ and $U=\{1,4,2,10\}$ from Fig. 9 (a)-(d), respectively.}}
    \Rev{
    \begin{tabular}{lcc}
\hline DPD Scheme & SLL [dB] & Main Lobe [dB] \\ 
\hline
Linear PA & $-[13.6, 7.8, 13.4, 12.5]$ & $[58.5, 32.2, 55.4, 48.0]$ \\
 w/o DPD  & $-[13.4, 7.7, 13.3, 12.4]$ & $[57.4, 31.1, 54.3, 46.9]$ \\
\shortstack{TD-GMP,$R=4$} & $-[13.5, 7.6, 13.2, 12.3]$ & $[58.5, 32.1, 55.4, 48.0]$  \\
FD-GMP  & $-[13.4, 7.6, 13.2, 12.3]$ & $[58.3, 32.0, 55.2, 47.7]$ \\
FD-NN  & $-[13.6, 7.6, 13.2, 12.3]$ & $[58.3, 32.1, 55.3, 47.8]$\\
FD-CNN   & $-[13.5, 7.6, 13.3, 12.3]$ & $[58.3, 32.1, 55.3, 47.8]$\\
\hline
\end{tabular}
}
    \label{tab:SLL_allband}
\end{table}
\subsubsection{Line-of-sight scenario}

Fig.~\ref{fig:BPs}~(a-d) show the beampatterns from a large linear array with $B=100$ antennas and without crosstalk in a \ac{los} scenario for different number of UEs, $U=\{1,2,4,10\}$, respectively. With a fixed total transmit power of the large array, PAs are operated in the same power level $P_{\text{PA}}=20$ dBm, which means each UE receives less power as $U$ grows.  The beampatterns from a SISO system are also shown in light blue solid and dashed lines for radiated in-band and \ac{oob} power, respectively. It is shown that \ac{oob} radiation can be beamformed strongly in certain directions, but also it can be very weak in certain directions or even become isotropic. The corresponding EVM and TRP ACLR results are shown in Table~\ref{tab:EVM_ACLR_LOS}.

Specifically, Fig.~\ref{fig:BPs}(a) and (b) illustrate the beampatterns for scenarios with $1$ and $4$ UEs, where in the latter scenario, the majority of power is directed towards the farthest UE to maintain uniform service quality. In the case of $U=1$,  distortion radiation is beamformed similarly to the in-band signal, achieving almost the same array gain, while the distortion in all the other directions remains minimal. Although TD-GMP yields better linearization  (with EVM and TRP-ACLR gain shown in Table~\ref{tab:EVM_ACLR_LOS}) over all FD-DPD schemes, it requires $61\times$, $54\times$, and $220\times$ more \acp{flop} than FD-GMP, FD-NN, and the proposed FD-CNN, respectively, as shown in Fig.~\ref{fig:FLOPs_vs_B}. All FD-DPDs meet the 3GPP requirements of EVM and TRP-ACLR. Similar trends are observed in Fig~\ref{fig:BPs}(b) for the case of $4$ UEs.  Most of the power and the OOB radiation are directed in the farthest UE direction, while much less OOB radiation is beamformed to other UEs' directions. Hence, FD-DPDs prove to be adequate sufficient to meet both EVM and ACLR requirements while substantially reducing computational complexity, whether dealing with single-user or multi-user cases with a single dominant UE. Notably, the proposed FD-CNN entails the least computational overhead.

Figure~\ref{fig:BPs}(c) shows the beampatterns for a LOS scenario with 2 UEs, each 28m from the BS and allocated equal power. The distortion radiation is strong in four directions, corresponding to the 2 UEs and 2 IMD beams. TD-GMP, unaffected by beamforming directions, delivers excellent linearization performance across all four directions, achieving the best EVM and TRP-ACLR (Table~\ref{tab:EVM_ACLR_LOS}), but requires 30 times, 27 times, and 110 times more FLOPs than FD-GMP, FD-NN, and FD-CNN, respectively. All three FD-DPDs achieve similar, though slightly inferior, in-band and OOB linearization across these directions, with extra beams generated and beamformed toward the two IMD directions. Despite the additional computational complexity from extra precoding, FD-DPDs remain preferable to TD-DPDs for scenarios with a low number of UEs, offering fewer FLOPs and satisfactory linearization per 3GPP requirements. However, the complexity advantage diminishes as the number of UEs and corresponding IMD beams increases.

Fig.~\ref{fig:BPs}(d) shows the beampatterns for 10 UEs, each allocated equal power. The OOB radiation becomes isotropic as the number of IMD beams ($\approx$ 1000) far exceeds the number of antennas (B=100), and is significantly weaker than in a SISO system (approximately 9 dB less), thanks to the array gain of 10 dB.
TD-GMP DPD provides the best linearization performance, unaffected by the large number of UEs and IMD beams. In contrast, the three FD-DPDs have limited linearization performance due to numerous IMDs, but still meet ACLR requirements due to the 9 dB OOB requirement relaxation from the array gain. Notably, FD-CNN requires approximately 20 times fewer FLOPs than TD-GMP (Fig.~\ref{fig:FLOPs_vs_U}).
However, the OOB requirement relaxation depends heavily on array gain, and with an increasing number of UEs, FD-DPDs may not meet OOB linearization requirements, making TD-DPDs the preferred choice. \Rev{Additionally, Table~\ref{tab:SLL_allband} shows that using DPDs has negligible impact on the SLL across different UEs, yielding SLL values similar to those without DPD and with linear PAs.}

\begin{table}[!t]
    \centering
    \caption{\Rev{EVM and TRP-ACLR results for DPD schemes in LOS channel with $B=100$, $U=1$, $N_{\text{d}}=64$, and $R=4$, with and without $-10$ dB linear and nonlinear crosstalk.}}
    \Rev{
    \begin{tabular}{lcc}
\hline DPD Scheme & \shortstack{$\text{ACLR}_{\text{TRP}}$ (dBc)\\ w/o \& w/ crosstalk} & \shortstack{$\text{EVM}_{\text{}}$ (\%)\\ w/o \& w/ crosstalk}  \\ 
\hline
 w/o DPD  & $[33.5, 32.5]$ & $[5.1, 11.2]$ \\
TD-GMP, $R=4$~\cite{GMP_2006} & $[48.2, 31.5]$ & $[0.6, 4.7]$  \\
FD-GMP~\cite{brihuega2022beam}  & $[38.2, 36.1]$  & $[1.5, 2.2]$ \\
FD-NN~\cite{tarver2021virtual}  & $[37.1, 35.0]$  & $[1.9, 2.4]$\\
FD-CNN   & $[37.2, 35.3]$  & $[2.0, 2.7]$\\
\hline
\end{tabular}
}
    \label{tab:EVM_ACLR_LOS_crosstalk}
\end{table}
\vspace{-0.2cm}

\begin{figure}[t]
	\centering
		\vspace*{0. \baselineskip}
	 \input{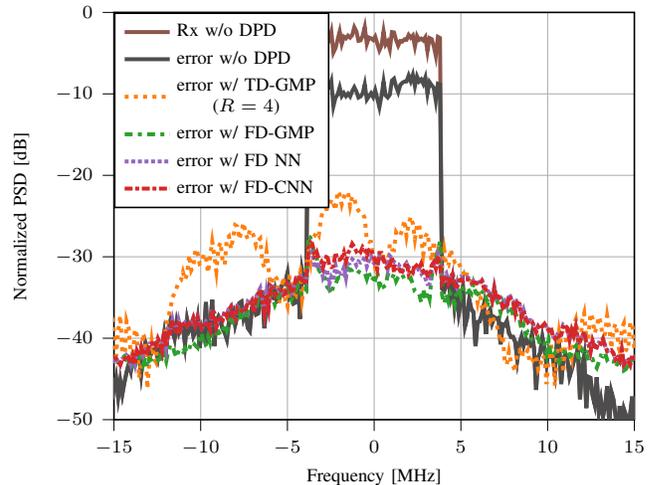}
			\vspace*{-0.1cm}
	\caption{\Rev{Normalized \ac{psd} and \ac{psd} of error at an intended UE w/o DPD and with different DPD schemes for $B=100$, $U=1$ in a LOS channel with $-10$ dB linear and nonlinear crosstalk.}}
	\label{fig:OTA_PSD_B100U1_crosstalk}
	\vspace*{-0.0cm}
\end{figure}

\begin{figure}
    \centering
    \input{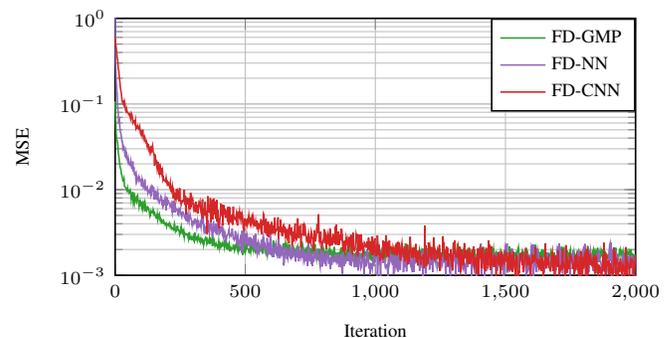}
    \caption{\RevMaj{Training loss of three FD-DPDs using the proposed FD-DPD learning algorithm for $B=100$, $U=1$ in a LoS channel.}}
    \label{fig:training_MSE}
\end{figure}

\Rev{
\vspace{-2mm}\subsubsection{Results With Crosstalk}
We analyze the impact of crosstalk on linearization performance. Specifically, we set both the PA input crosstalk level and the antenna crosstalk level to $-10$ dB. Fig.~\ref{fig:OTA_PSD_B100U1_crosstalk} shows the OTA \ac{psd} under crosstalk with $B=100$ BS antennas and $U=1$ UE in a LOS scenario. The corresponding EVM and ACLR results, including those without crosstalk, are summarized in Table~\ref{tab:EVM_ACLR_LOS_crosstalk}. The results show that TD-GMP DPD performance is significantly degraded by crosstalk due to ILA, whereas the three FD-DPDs experience less degradation and achieve better in-band linearization, despite only slight improvement in OOB radiation. The results suggest that FD-DPDs can effectively mitigate PA and antenna crosstalk using the proposed learning approach in Algorithm~\ref{alg:FD_DPD_learn}. To reduce the impact of crosstalk, TD-DPD needs to either change the optimization algorithm to an iterative closed-loop algorithm~\cite{brihuega2020digital} or utilize specific and more complex processing models such as the dual-input DPD model~\cite{hausmair2017digital}.
}
\RevMaj{
\subsubsection{Convergence Speed}
We analyze the convergence speed of the training process for the three FD-DPD schemes using the proposed learning algorithm in Algorithm~\ref{alg:FD_DPD_learn}, within a LoS scenario involving $100$ antennas and $1$ UE. Fig.~\ref{fig:training_MSE} illustrates the training loss as a function of the number of iterations. The results show that the proposed learning algorithm effectively optimizes all three FD-DPD schemes. FD-GMP converges faster due to its simpler model with fewer parameters, while FD-CNN converges more slowly, likely due to the complexity of the convolutional layers compared to the use of IDFT and FFT operations in FD-GMP and FD-NN.
}

\section{Conclusion}
In this study, we analyzed complexity-performance trade-offs of TD and FD DPD methods in fully digital MIMO systems within \ac{fr1}, focusing on linearization requirements in various channel conditions, UE numbers, and antenna crosstalk. \RevMaj{Our findings suggest that lower-complexity FD-DPDs can be used in many MIMO scenarios, thanks to the relaxed OOB linearization requirements. We proposed a novel FD CNN-based DPD with lower complexity than other TD and FD benchmarks when the number of BS antennas exceeds 4 or UEs are fewer than 64. Additionally, we proposed a new learning algorithm for any FD-DPD with a differentiable structure. We conducted a comprehensive analysis across LOS and isotropic scattering channels, with varying number of UEs in \ac{fr1}.  Our results show that FD-DPD, particularly our FD-CNN DPD, perform effective in both channels, when dealing with fewer UEs, satisfying 3GPP in-band and OOB linearity requirements while reducing complexity. However, in scenarios with a high number of UEs ($>10$ in LOS and $>4$ in isotropic scattering channels), TD-DPDs are preferable to meet the 3GPP requirements. Furthermore, the proposed learning algorithm for FD-DPDs achieves better linearization performance than TD-DPDs using ILA under antenna crosstalk. These insights provides valuable guidance for selecting DPD schemes tailored to specific complexity constraints and linearization requirements in fully digital massive MU-MIMO systems in FR1. The complexity reduction achieved with the FD-CNN model also encourages the use of digital arrays at higher frequencies, such as FR2.}

\appendix

\section{Computational Complexity Calculation}\label{appdx:complexity}
\RevMaj
{
\vspace{-0mm}\subsection{TD-GMP}\label{appdx:complexity_TD_GMP}
The number of \acp{flop} required for the GMP with each input sample is computed as~\cite[Eq. (16-18)]{complexityGMP}
\begin{align}
C^{\text{TD-GMP}}_{\text{Samp}}= &8\big((M_{\text{TD}}+1)(K+2 K G)-\frac{G(G+1)}{2}(K-1)\big) 
\nonumber \\
+10&+ 2K+2(K-1) G+2 K \min (G, M_{\text{TD}}). \label{eq:GMP_FLOPs_per_sample}
\end{align}
Each TD GMP-based DPD corresponds to an antenna of the massive MU-MIMO system. The number of FLOPs required per QAM symbol can be calculated:
\begin{align}
   C^{\text{TD-GMP}}_{\text{}} = C^{\text{TD-GMP}}_{\text{Samp}} RN_{\text{d}}B/N_{\text{d}}\,=C^{\text{TD-GMP}}_{\text{Samp}} RB\,.
   \label{eq:TD_GMP_Complexity}
\end{align}
\subsection{FD-GMP}\label{appdx:complexity_FD_GMP}
In the FD-GMP~\cite{brihuega2021frequency,brihuega2022beam}, the additional IDFTs and DFTs introduce an extra complexity cost. More importantly, the OOB subcarriers of the predistorted signal are no longer empty after the DFTs, which leads to an additional complexity cost associated with the precoding. The number of \acp{flop} required per QAM symbol is calculated as
\begin{align}
    &C^{\text{FD-GMP}} = \frac{1}{N_{\text{d}}}\Big(\underbrace{C^{\text{TD-GMP}}_{\text{Samp}}N_{\text{d}}R(U+V)}_{\text{GMP of Multiple beams}} \nonumber \\& +  \underbrace{\big(U+(U+V)KM_{\text{FD}} \big)(4N\log_2N -6N+8)}_{\text{(I)DFTs}} \nonumber\\
    &+ \underbrace{(6+2)(UN_{\text{g}} + NV)B}_{\text{Extra precoding}} \Big) \label{eq:FD_GMP_Complexity} \\
    & \approx \left( C^{\text{TD-GMP}}_{\text{Samp}} + 4KM_{\text{FD}}\log_2N + 8 B\right)R(U+V), \label{eq:FD_GMP_Complexity_Approx}
\end{align}
where the factors $6$ and $2$ in the extra precoding part of~\eqref{eq:FD_GMP_Complexity} are for complex-number multiplication and addition, respectively. Considering the \ac{fft}, the complexity of a $N$-size (I)DFT requires $(N\log_2N-3N+4)$ real multiplications and $(3N\log_2N -3N+4)$ real additions~\cite{sorensen1986computing}, and there are $U$ IDFTs and $(U+V)KM$ DFTs. 
\subsection{FD-NN}\label{appdx:complexity_FD_NN}
Similar to FD-GMP, FD-NN requires additional \acp{idft}, \acp{dft}, and precodings. For each UE, the number of \acp{flop} required for the FD-NN per QAM symbol can be computed as
\begin{align}
        &C_{\text{}}^{\text{FD-NN}}  =  \Big( \underbrace{N\big(4U (M_{\text{FD}}+1) D  + 2(K^{\text{NN}}-1) D^2 + 4DU\big)}_{\text{NN}} \notag \\ &+  \underbrace{2U(4N\log_2N -6N+8))}_{\text{(I)DFTs}} + \underbrace{(6+2)N_gUB}_{\text{Extra precoding}} \Big) /N_{\text{d}} \\
         & \quad \quad \quad \approx (D+D^2/U+8B)RU.  \label{eq:FD_NN_Complexity}
\end{align}
}
\subsection{FD-CNN}\label{appdx:complexity_FD_CNN}
To calculate the complexity of the convolution layer part, we note that each 2D convolution operation with a kernel size of $K_{\text{C}}$  requires $(2K_{\text{C}}^2 -1)$ \acp{flop}, which consists of $K_{\text{C}}^2$ real-valued multiplications and $(K_{\text{C}}^2-1)$ real-valued additions. For the first layer of the FD-CNN, $N^{\text{Conv1}}$ convolutional kernels operate on $2U$ real-valued UE symbol matrices with stride length $K_{\text{S}}$, which contributes to the number of \acp{flop} for the first convolution layer part as
$
	C_{\text{Conv1}}^{\text{FD-CNN}} = 2N^{\text{Conv}}U (2K_{\text{C}}^2 -1) \left({\lceil\sqrt{N_{\text{d}}}\rceil}/{K_{\text{S}}}\right)^2. 
$
Similarly, we can calculate the number of \acp{flop} for the second layer of the FD-CNN consisting of $2U$ convolutional kernels, which gives the same complexity as the first layer:  $C_{\text{Conv2}}^{\text{FD-CNN}} = C_{\text{Conv1}}^{\text{FD-CNN}}$. The number of FLOPs for the fully-connected layer part is
$
	C_{\text{FC}}^{\text{FD-CNN}} =  8N_{\text{d}}U \left({\lceil\sqrt{N_{\text{d}}}\rceil}/{K_{\text{S}}}\right)^2. \label{eq:FD_CNN_Complexity_FC}
$
In total, the number of \acp{flop} required for the FD-CNN per UE and QAM symbol is calculated as
\begin{align}
C^{\text{FD-CNN}} & = (C_{\text{Conv1}}^{\text{FD-CNN}} + C_{\text{Conv2}}^{\text{FD-CNN}} + C_{\text{FC}}^{\text{FD-CNN}})/N_{\text{d}}  \\
&\approx  8(K_{\text{C}}^2 N^{\text{Conv1}} + N^{\text{d}})U.
	\label{eq:FD_CNN_Complexity}
\end{align}

\balance

\bibliographystyle{IEEEtran}

\bibliography{./bibliography/IEEEabrv,reference_list}
\end{document}